\documentclass[usenatbib]{mn2e}
\usepackage{epsf}
\usepackage{amsmath,amssymb}
\usepackage{epsfig}
\usepackage{psfrag}
\usepackage{color}

\DeclareFontFamily{U}{euc}{}
\DeclareFontShape{U}{euc}{m}{n}{<-6>eurm5<6-8>eurm7<8->eurm10}{}%
\DeclareSymbolFont{AMSc}{U}{euc}{m}{n} 
\DeclareMathSymbol{\upzeta}{\mathord}{AMSc}{"10}

\newcommand{\bfm}[1]{\mbox{\boldmath $#1$}}
\newcommand{\ie}{\rm i.e.}
\newcommand{\etal}{\rm et al.}
\newcommand{\eg}{\rm e.g.}
\newcommand{\divb}[1]{\mbox{${\bfm \nabla}\cdot {\bf B}=0$}\ }
\newcommand{\beq}{\begin{equation}}
\newcommand{\eeq}{\end{equation}}

\newcommand{\tr}{\mbox{tr}}
\newcommand{\half}{\frac{1}{2}}
\newcommand{\n}{\noindent}
\newcommand{\bec}{\begin{center}}
\newcommand{\eec}{\end{center}}

\newcommand{\tbr}{\bar{\tau}}
\newcommand{\ptwo}[2]{\frac{{\partial}^2 #1}{\partial {#2}^2}}
\newcommand{\p}[2]{\frac{{\partial} #1}{\partial {#2}}}
\newcommand{\mat}[1]{\mbox{\boldmath $\mathsf{#1}$}}
\newcommand{\mx}[1]{{#1}_{\mbox{\scriptsize{max}}}}

\newcommand{\lesim}{\,\raisebox{-0.4ex}{$\stackrel{<}{\scriptstyle\sim}$}\,}

\title[Modelling weakly ionized plasmas]{A three-dimensional numerical method for modelling weakly ionized plasmas}

\author[S.\ O'Sullivan and T.P.\ Downes]
{Stephen O'Sullivan$^{1}$ \thanks{E-mail: stephen.osullivan@ucd.ie
(SOS); turlough.downes@dcu.ie (TPD)} and Turlough P. Downes$^{2}$ ${}^{\star}$\\
$^{1}$UCD School of Mathematical Sciences, University College Dublin, Belfield, Dublin 4, Ireland\\
$^{2}$School of Mathematical Sciences and National Centre for Plasma Science and Technology,\\ 
Dublin City University, Glasnevin, Dublin 9, Ireland
}
\begin{document}

\date{Accepted 2006 December 19. Received 2006 December 13; in original form 2006 November 17}

\pagerange{\pageref{firstpage}--\pageref{lastpage}}
\pubyear{2006}

\maketitle

\label{firstpage}

\begin{abstract}

Astrophysical fluids under the influence of magnetic fields are often subjected to single-fluid or two-fluid approximations. In the case of weakly ionized plasmas however, this can be inappropriate due to distinct responses from the multiple constituent species to both collisional and non-collisional forces. As a result, in dense molecular clouds and proto-stellar accretion discs for instance, the conductivity of the plasma may be highly anisotropic leading to phenomena such as Hall and ambipolar diffusion strongly influencing the dynamics.

Diffusive processes are known to restrict the stability of conventional numerical schemes which are not implicit in nature. Furthermore, recent work establishes that a large Hall term can impose an additional severe stability limit on standard explicit schemes. Following a previous paper which presented the one-dimensional case, we describe a fully three-dimensional method which relaxes the normal restrictions on explicit schemes for multifluid processes. This is achieved by applying the little known Super TimeStepping technique to the symmetric (ambipolar) component of the evolution operator for the magnetic field in the local plasma rest-frame, and the new Hall Diffusion Scheme to the skew-symmetric (Hall) component.

\end{abstract}


\section{Introduction}
\label{introduction}

Numerical schemes used in simulations of astrophysical plasmas are frequently  derived from single-fluid magnetohydrodynamic (MHD) models\footnote{We associate the multiplicity of the fluids described by a model to the number of fluids treated distinctly in the derived numerical scheme.}. The most common example of this is ideal MHD, with assumptions including infinite conductivity and negligible Hall current. Extended models within the single-fluid framework are commonly used for finite scalar conductivity and Hall current. Furthermore, two-fluid models are used when the drift of a neutral component through the bulk plasma is considered important. With reference to the generalized Ohm's law, we now briefly survey the physical motivations for departing from models based on ideal MHD. The discussion makes a progression through various models arriving at the argument for a fully multifluid numerical approach to weakly ionized plasmas. 


The generalized Ohm's law for collisional gases describes the dependencies of electric currents on the relative drift of charged particles due to effects both mediated by, and independent of, magnetic fields. In the latter case, for example, electron pressure can cause electrons in a local condensation of gas to diffuse more quickly than ions due to greater thermal velocities. The resulting separation of charge creates an electric force coupling the ion and electron gases in a process known in plasma physics as ambipolar diffusion \citep{cowling}. In the following however, electron pressure is neglected under the assumptions that $L\gg c/\omega_{\rm pe}$ and $L\gg r_{\rm e}$ where $L$ is the scale length of the plasma, $\omega_{\rm pe}$ is the electron plasma frequency, and $r_{\rm e}$ is the electron gyroradius. The term \emph{ambipolar diffusion} is now used without ambiguity to describe an entirely different, magnetically mediated phenomenon of neutral drift, as more commonly discussed in astrophysical contexts (\citealt{mestel56, spitzer78}; and more recently, \citealt{wardleng99}).

Defining $\bfm{E}'$ as the electric field in the local rest frame of the bulk plasma, and considering only effects dependent on the presence of a magnetic field, the generalized Ohm's law can be written as

\begin{eqnarray}
\bfm{E}'&=& \bfm{\sigma}^{-1}\cdot\bfm{J} \nonumber \\ 
        &=& r_{\rm O} \bfm{J}_\parallel +r_{\rm H}\bfm{J}_\perp\times\hat{\bfm{B}} +r_{\rm A}\bfm{J}_\perp .
\label {eqn-ohm}
\end{eqnarray}

\n In this equation $\bfm{\sigma}$ is the tensor conductivity of the plasma, and $r_{\rm O}$, $r_{\rm H}$, $r_{\rm A}$ are the corresponding Ohmic (field-parallel), Hall, and ambipolar (Pedersen) resistivities respectively. The explicit form of the conductivity for a weakly ionized plasma will be discussed in section~\ref{multifluid_equations}, however, it is worth pointing out some general properties of equation~\eqref{eqn-ohm} before proceeding.

While collisions may produce rich and complex physics via their influence on currents, Hall diffusion can operate independently of collisional forces. (Note that we refer to the Hall term as diffusive in the sense that it contributes to the violation of field freezing, however, it is dispersive in nature and twists, rather than diffuses, the magnetic field.) Considering firstly the special case of fully ionized gases where \mbox{$r_{\rm A}= r_{\rm O}\equiv r_{\rm res}$}, the Ohmic and ambipolar terms in equation~\eqref{eqn-ohm} may be combined into a single resistive term $r_{\rm res}\bfm{J}$.  For $L\lesim c/\omega_{\rm pi}$, where $\omega_{\rm pi}$ is the ion cyclotron frequency, the greater inertia of the ions causes them to decouple from the electrons (even when collisions are unimportant and $r_{\rm res}\to 0$) and the Hall term $r_{\rm H}\bfm{J}_\perp\times\hat{\bfm{B}}$ in equation~\eqref{eqn-ohm} becomes significant. This regime is frequently approximated via the single-fluid Hall-MHD model~\citep[see, for example, ][]{huba,mininni05}.

Furthermore, when collisions are important, disparate resistive effects impede the flows of currents in senses both parallel and perpendicular to the magnetic field. In fully ionized plasmas, or weakly ionized plasmas where magnetic forces on the charged species are dominated by collisional drag on the neutrals, the electron drift with respect to the bulk plasma is fully determined by the electric current and a single-fluid model is tenable. Moreover, if the Hall effect is negligible ($L\gg c/\omega_{\rm pi}$), so-called resistive MHD is retrieved with a scalar conductivity \mbox{$\sigma_{\rm res}=r_{\rm res}^{-1}$} and corresponding Ohm's law \mbox{$\bfm{E}'=r_{\rm res}\bfm{J}$}.

In incompletely ionized plasmas when magnetic forces on the charged species dominate collisional drag, ambipolar diffusion occurs as the charged particles remain tightly coupled to the magnetic field while drifting through the neutral gas. Under these conditions it may be appropriate to use two fluid models which represent the plasma as an ion gas interacting with a neutral component \citep{draine80, toth94, sm97, stone97}.  

Recently, \cite{pandey} have asserted that in weakly ionized plasmas, collisional coupling with the neutrals reduces the effective gyrofrequency of ions by a factor $\rho_{\rm i}/\rho$, where $\rho_{\rm i}$ is the ion gas density and $\rho$ is the bulk plasma density. The Hall effect then becomes significant under the relaxed condition \mbox{$L\lesim  \rho c/\rho_{\rm i} \omega_{\rm pi}$}. Additionally, given the potential importance of charge-carrying grain species in molecular clouds \citep[\eg][]{wardle98, wardle04, cr02, falle03}, it is clear that a genuinely multifluid approach may often be necessary to capture the complex interplay of resistive effects due to  relative motions between species. Similarly, the conditions in proto-stellar accretion discs may warrant a multifluid treatment \citep[\eg][]{ss02a, ss02b, wardle04, salmeron}.

The numerical difficulties introduced by the presence of significant Hall diffusion have 
been outlined by F03, and \cite{osd06}.  Both of these works put forward one-dimensional numerical methods for 
multifluid MHD of weakly ionized plasmas which overcome these 
difficulties.  However, the method presented in Paper~I has the significant 
advantage of being explicit and hence being comparatively easy to implement, 
particularly in codes employing techniques crucial to large scale simulations 
such as parallel domain decomposition and adaptive mesh refinement (AMR).

In this paper we present the extension of the method described in Paper~I to three 
dimensions.  Section~\ref{multifluid_equations} details the 
multifluid equations governing weakly ionized plasmas.  In section
\ref{numerical_method} we discuss the numerical method used to integrate 
these equations, focusing section~\ref{sec_diff} on the treatment of 
magnetic diffusion with particular emphasis on Hall diffusion.  In 
section \ref{numerical_tests} we present three-dimensional results of 
shock-tube tests and simulations of three-dimensional turbulence in both 
ambipolar and Hall diffusion regimes.  Finally, in section~\ref{conclusions} 
we make some concluding remarks.

\section{The multifluid equations}
\label{multifluid_equations}

We assume a weakly ionized plasma such that the mass density is dominated by the neutral component of the gas.  Then, relative to the scale length of the system, if particles of a given charged species have small mean free paths in the neutral gas, or small Larmor radii, their pressure and inertia may be neglected. See F03 for a more detailed discussion.


For convenience it is assumed that there is no mass transfer between species.  It is straightforward, however, to insert the necessary terms for a more general treatment to include mass transfer if necessary.  The equations governing the evolution of the weakly ionized plasma can then be written as 

\beq
\frac{\partial \rho_n}{\partial t} + \nabla\cdot \left(\rho_n
	\bfm{q}_n\right)  = 0 ,
\label{mass}
\eeq

\beq
\frac{\partial \rho_1 \bfm{q}_1}{\partial t} 
+ \nabla\cdot\left( \rho_1 \bfm{q}_1 \bfm{q}_1 + p_1\mat{I}\right)  =  \bfm{J}\times\bfm{B} ,
\label{neutral_mom}
\eeq

\beq
\frac{\partial e_1}{\partial t} + \nabla\cdot
	\left[\left(e_1+p_1\right)\bfm{q}_1\right] =\bfm{J}\cdot\bfm{E}  +\sum_{n=1}^N H_n  ,
	\label{neutral_en} 
\eeq

\beq
\frac{\partial \bfm{B}}{\partial t} +\nabla\cdot(\bfm{q}_1\bfm{B}-\bfm{B}\bfm{q}_1) = -\nabla\times\bfm{E}' ,
\label{B_eqn} 
\eeq

\beq
\alpha_n \rho_n\left(\bfm{E} + \bfm{q}_n \times \bfm{B} \right) +
\rho_n \rho_1 K_{n\,1}(\bfm{q}_1-\bfm{q}_n) =  0 , 
\label{charged_mom}
\eeq

\beq
H_n +G_{n\,1} +\alpha_n \rho_n \bfm{q}_n \cdot \bfm{E} =0 ,
\label{charged_en}
\eeq

\beq
\nabla\cdot\bfm{B}=0 ,
\label{divB}
\eeq

\beq
\bfm{J}=\nabla\times\bfm{B} ,
\label{eqn-J}
\eeq

\beq
\sum_{n=2}^N \alpha_n \rho_n  =  0 ,
 \label{charge_neutrality}
\eeq

\beq
\sum_{n=2}^N \alpha_n \rho_n\bfm{q}_n  =  \bfm{J} .
 \label{current}
\eeq

\noindent In the preceding equations, the subscripts denote the species, with a subscript of 1
indicating the neutral fluid.  The variables $\rho_n$, $\bfm{q}_n \equiv
(u_n,\,v_n,\,w_n)^{\rm T}$, $p_n$ and $e_n$ are the mass
density, velocity, pressure and total energy respectively of species~$n$.  In general we assume a closure relation

\beq 
e_n=\frac{p_n}{\gamma_n-1}+\half\rho_n q_n^2  ,
\label{eqn-eos}
\eeq

\n where $\gamma_n$ is the ratio of specific heats for species~$n$. However, for the test cases described here, an isothermal equation of state is assumed allowing us to disregard equations~\eqref{neutral_en}~and~\eqref{charged_en} and use the closure relation 

\beq
a^2=p_1/\rho_1 ,
\label{eqn-eosiso}
\eeq

\n where $a$ is the (constant) isothermal soundspeed.
The identity matrix, current density and
magnetic flux density are represented by $\mat{I}$, $\bfm{J}$, $\bfm{B}$ respectively.  $\bfm{E}'$ is related to the full electric field $\bfm{E}$ by 

\beq
\bfm{E}=-\bfm{q}_1\times\bfm{B} +\bfm{E}' .
\label{eqn-efield}
\eeq
Additionally, with reference to species~$n$: $K_{n\,1}$ describes the collisional interaction with the neutral fluid, $\alpha_n$ is
the charge-to-mass ratio, $G_{n\,1}$ is the energy
transfer rate to the neutral fluid, and $H_n$ is the energy source or
sink.  Note that in
general $K_{n\,1}$ and $G_{n\,1}$ may depend on the temperatures and
relative velocities of the interacting species.  Equations~\eqref{mass}~to~\eqref{charged_en} are derived from the
conservation equations for mass (of all species), neutral species momentum, 
neutral species energy, magnetic flux, charged species momentum, and charged species energy respectively.  Equations~\eqref{divB}~to~\eqref{current} describe the solenoidal condition, Amp\`ere's law (with displacement current neglected), charge neutrality, and charge current respectively. We refer the reader to F03 and
\cite{cr02} for a more detailed discussion.

For a weakly ionized plasma, the generalized Ohm's law can be written in terms of contributions from Ohmic, Hall, and ambipolar terms \citep[\eg][]{wardleng99} as

\beq
\bfm{E}' = \bfm{E}_{\rm O} +\bfm{E}_{\rm H} +\bfm{E}_{\rm A} ,
\label{eqn_E1}
\eeq

\n where

\beq
\bfm{E}_{\rm O}=(\bfm{J}\cdot\bfm{a}_{\rm O})\bfm{a}_{\rm O} ,
\label{eqn_EO1}
\eeq

\beq
\bfm{E}_{\rm H}=\bfm{J}\times\bfm{a}_{\rm H} ,
\label{eqn_EH1}
\eeq

\beq
\bfm{E}_{\rm A}=-(\bfm{J}\times\bfm{a}_{\rm A})\times\bfm{a}_{\rm A} .
\label{eqn_EA1}
\eeq

\n We use the definition

\beq
\bfm{a}_{\rm X}  \equiv  f_{\rm X} \bfm{B} ,
\eeq

\n where ${\rm X}$ is one of ${\rm O}$, ${\rm H}$, or ${\rm A}$ and

\begin{eqnarray}
f_{\rm O} & = & \sqrt{r_{\rm O}}/B , \\
f_{\rm H} & = & r_{\rm H}/B , \\
f_{\rm A }& = & \sqrt{r_{\rm A}}/B .
\end{eqnarray}

\n Here $r_{\rm O}$, $r_{\rm H}$ and $r_{\rm A}$ are the Ohmic, Hall and ambipolar
resistivities respectively defined by the relations

\begin{eqnarray}
r_{\rm O} & = & \frac{1}{\sigma_{\rm O}} , \\
r_{\rm H} & = & \frac{\sigma_{\rm H}}{\sigma_{\rm H}^2 + \sigma_{\rm A}^2} , \\
r_{\rm A} & = & \frac{\sigma_{\rm A}}{\sigma_{\rm H}^2 + \sigma_{\rm A}^2} ,
\end{eqnarray}

\n with the conductivities given by

\begin{eqnarray}
\sigma_{\rm O} & = & \frac{1}{B}\sum_{n=2}^N \alpha_n \rho_n \beta_n , \\
\sigma_{\rm H} & = & \frac{1}{B}\sum_{n=2}^N \frac{\alpha_n \rho_n}{1+\beta_n^2} ,\\
\sigma_{\rm A} & = & \frac{1}{B}\sum_{n=2}^N \frac{\alpha_n \rho_n \beta_n}{1+\beta_n^2} ,
\end{eqnarray}

\noindent The Hall parameter $\beta_n$ for species~$n$ is 

\begin{equation}
\beta_n = \frac{\alpha_n B}{K_{1\,n}\rho_1} .
\label{eqn-hallpar}
\end{equation}

\section{Numerical method}
\label{numerical_method}

We assume a piecewise constant solution on a uniform mesh
of spacing $h$ in each of the $x$, $y$ and $z$ directions. If the solution has been marched forward in time through $l$ (not necessarily uniform) intervals, we denote the current time as $t^l$ and seek the solution at some later time \mbox{$t^{l+1}\equiv t^l+\tau$}. Cell \mbox{$(i,\,j,\,k)$} of the mesh is defined as the volume $\{(x,y,z): (i-1/2)h\le x\le (i+1/2)h, (j-1/2)h\le y\le (j+1/2)h, (k-1/2)h\le z\le (k+1/2)h\}$. Then given any quantity $D(x,\,y,\,z,\,t)$ continuously defined on the mesh volume, the average value over the cell $\mbox{(i,\,j,\,k)}$ at time $t^l$, is denoted by $D_{i,\,j,\,k}^l$ and defined at the cell centre. Note that for the sake of clarity we may drop any of the indices~$i$,~$j$,~$k$,~or~$l$ if no ambiguity arises.

To obtain the full solution at time $t^{l+1}$, standard finite volume
integration methods are applied to all terms in the partial differential equations~\eqref{mass}~to~\eqref{B_eqn} with the exception of the diffusive term $-\nabla\times\bfm{E}'$ on the right hand side of equation~\eqref{B_eqn} which we discuss in the next section. The time integration is
multiplicatively operator split with each operation carried out to
second order spatial and temporal accuracy in a straightforward extension of the methods described in Paper~I. Overall second order accuracy in time is maintained by permuting the order of operations \citep{strang}. Charged species velocities and pressures may be derived algebraically by means of equations~\eqref{charged_mom}~and~\eqref{charged_en}; the approach to the charged velocity is described in appendix~\ref{charged_vel_app}. Finally, the \mbox{$\nabla\cdot\bfm{B}=0$} constraint is applied during each timestep. Further discussion of this is deferred to section~\ref{sec-divb}.

\subsection{Treatment of magnetic diffusion}
\label{sec_diff}

We now focus on the numerical methods for integration of the magnetic diffusion terms. The induction equation without the hyperbolic terms is

\begin{eqnarray}
\frac{\partial \bfm{B}}{\partial t} &=&  -\nabla\times\bfm{E}' \nonumber \\
        &=& -\nabla\times(\bfm{E}_{\rm O} +\bfm{E}_{\rm H} +\bfm{E}_{\rm A}) ,
\label{B_eqn3} 
\end{eqnarray}

\n using equation~\eqref{eqn_E1}. To proceed, we carry out the expansions

\beq
\nabla\times\bfm{E}_{\rm X}=  \bfm{F}^1_{\rm X} +\bfm{F}^2_{\rm X} ,
\label{eqn-Bexp}
\eeq

\n where the subscript ${\rm X}$ is one of ${\rm O}$, ${\rm H}$, or ${\rm A}$. The corresponding linear and second order terms, $\bfm{F}^1_{\rm X}$ and $\bfm{F}^2_{\rm X}$ respectively, are

\bec
\begin{eqnarray}
\bfm{F}^1_{\rm O}= && -[\bfm{a}_{\rm O}\cdot(\nabla\times\bfm{J})]\bfm{a}_{\rm O} +[(\bfm{a}_{\rm O}\cdot\nabla)\bfm{J})]\times\bfm{a}_{\rm O} \nonumber \\
                   && +a_{\rm O}^2 \nabla\times\bfm{J} ,
\label{eqn_FO1}
\end{eqnarray}
\eec

\bec
\begin{eqnarray}
\bfm{F}^2_{\rm O}=	&& -[\bfm{a}_{\rm O}\cdot(\nabla\times\bfm{a}_{\rm O})]\bfm{J} +[(\bfm{J}\cdot\nabla)\bfm{a}_{\rm O}]\times\bfm{a}_{\rm O}  \nonumber \\
                        && +2(\bfm{J}\cdot\bfm{a}_{\rm O})[\nabla\times\bfm{a}_{\rm O}] ,
\label{eqn_FO2}
\end{eqnarray}
\eec

\bec
\begin{eqnarray}
\bfm{F}^1_{\rm H}= &&    ~~(\bfm{a}_{\rm H}\cdot\nabla)\bfm{J} ,
\label{eqn_FH1}
\end{eqnarray}
\eec

\bec
\begin{eqnarray}
\bfm{F}^2_{\rm H}=	&& -   (\bfm{J}\cdot\nabla)\bfm{a}_{\rm H} +(\nabla\cdot\bfm{a}_{\rm H})\bfm{J} ,
\label{eqn_FH2}
\end{eqnarray}
\eec

\bec
\begin{eqnarray}
\bfm{F}^1_{\rm A}=  &&  ~~[\bfm{a}_{\rm A}\cdot(\nabla\times\bfm{J})]\bfm{a}_{\rm A} -[(\bfm{a}_{\rm A}\cdot\nabla)\bfm{J})]\times\bfm{a}_{\rm A}   ,
\label{eqn_FA1}
\end{eqnarray}
\eec

\bec
\begin{eqnarray}
\bfm{F}^2_{\rm A}= &&  +[\bfm{a}_{\rm A}\cdot(\nabla\times\bfm{a}_{\rm A})]\bfm{J} -[(\bfm{J}\cdot\nabla)\bfm{a}_{\rm A})]\times\bfm{a}_{\rm A}     \nonumber  \\ 
		      &&  -2(\bfm{J}\cdot\bfm{a}_{\rm A})[\nabla\times\bfm{a}_{\rm A}] +(\nabla a_{\rm A}^2)\times\bfm{J} .
\label{eqn_FA2}
\end{eqnarray}
\eec

In the following, we treat the discretization of equation~\eqref{B_eqn3} as a two part process. Firstly, under certain assumed conditions, the stability properties of schemes for the dominant terms are explored. Secondly, a correction must be made to the field updated through such a scheme to include any neglected small terms. The latter step is essential for consistency with the governing equation and is discussed in more detail in section~\ref{sec-correction}. However for now, we focus on the first step of the process.

Under the assumption of small perturbations in $\bfm{B}$ about a mean field, the second order terms $\bfm{F}^2_{\rm O}$, $\bfm{F}^2_{\rm H}$, and $\bfm{F}^2_{\rm A}$ are small in comparison with $\bfm{F}^1_{\rm O}$, $\bfm{F}^1_{\rm H}$, and $\bfm{F}^1_{\rm A}$ respectively. Additionally, under the often reasonable assumption that collisional drag on charged particles is dominated by magnetic forces, the Ohmic resistivity $r_{\rm O}$ is weak (F03) and hence $\bfm{F}^1_{\rm O}$ is also small. The stability of a scheme can then be investigated through analysis of the reduced induction equation

\beq
\frac{\partial \bfm{B}}{\partial t} \approx  \bfm{F}^1_{\rm H} +\bfm{F}^1_{\rm A} .
\label{B_eqn4} 
\eeq

The relative importance of the ambipolar and Hall resistivities may now be parametrized by \mbox{$\eta\equiv r_{\rm A}/|r_{\rm H}|$}. From this point, time intervals are normalized such that \mbox{$\tbr\equiv\tau/\tau^{\perp}$}, where $\tau^{\perp}$ is the characteristic cell crossing time for diffusion perpendicular to the magnetic field given by

\beq
\tau^{\perp}=\frac{h^2}{2\sqrt{r_{\rm H}^2+r_{\rm A}^2}} .
\eeq

\n Equation~\eqref{B_eqn4} can be rewritten as

\beq
\p{\bfm{B}}{t}=-\mat{G}\bfm{B} ,
\eeq

\n where, using \mbox{$\bfm{b}\equiv\bfm{B}/B$}, the matrix operator $\mat{G}$ is given by \mbox{$\mat{G}=\mat{G}_{\rm H}+\mat{G}_{\rm A}$} with

\bec
\begin{eqnarray}
\mat{G}_{\rm H}  = && -r_{\rm H}(\bfm{b}\cdot\nabla)(\nabla\times\cdot) ,
\end{eqnarray}
\eec

\bec
\begin{eqnarray}
\mat{G}_{\rm A} = && ~~r_{\rm A}[\bfm{b}\cdot(\nabla\times(\nabla\times\cdot))]\bfm{b}     \nonumber \\
          && -r_{\rm A}[(\bfm{b}\cdot\nabla)(\nabla\times\cdot)]\times\bfm{b} .
\end{eqnarray}
\eec

 The discretized form of the operator $\mat{G}$ at time level~$l$, denoted~$\mat{G}^l$, is obtained by using the second order derivative dicretizations

\beq
\left(\ptwo{B}{x}\right)_i=\frac{ B_{i+1} -2 B_{i} -B_{i-1}}{h^2} ,
\label{eqn-der1}
\eeq

\beq
\left(\frac{\partial^2 B}{\partial x\,\partial y}\right)_{i\,j}=\frac{ B_{i+1\,j+1} -B_{i+1\,j-1} -B_{i-1\,j+1} +B_{i-1\,j-1}}{4 h^2} ,
\label{eqn-der2}
\eeq

\n and similar expressions for other terms. Note that schemes with simpler discretizations and superior formal stability properties may be derived by replacing equation~\eqref{eqn-der1} with \mbox{$(\partial^2 B /\partial x^2)_i = ( B_{i+2} -2 B_{i} -B_{i-2})/4h^2$}. We do not consider such schemes further as they are odd-even decoupled and hence subject to instability.

For the purpose for stability analysis, we take a numerical wave of the form

\beq
\bfm{B}^l_{i\,j\,k}=\bfm{B}_0 \mbox{e}^{\mbox{i}\, \bfm{\omega}\cdot\bfm{i}} ,
\label{eqn-numwave}
\eeq

\n where $\bfm{B}_0$ is the wave amplitude, \mbox{$\mbox{i}\equiv\sqrt{-1}$}, \mbox{$\bfm{i}=(i,\,j,\,k)$}, and \mbox{$\bfm{\omega}=(\omega_x,\,\omega_y,\,\omega_z)$}. Second order derivatives of $\bfm{B}$ may now be replaced using 

\beq
\frac{\partial^2}{{\partial x^2}}\rightarrow \lambda_{x\,x}\equiv -2(1-\cos\omega_x) ,
\label{eqn_lxx}
\eeq

\beq
\frac{\partial^2}{{\partial x\,\partial y}}\rightarrow \lambda_{x\,y}\equiv-\sin\omega_x\sin\omega_y ,
\label{eqn_lxy}
\eeq

\n and similar substitutions for other terms. A matrix $\mat{\Lambda}$ can then be defined whose $(x,\,y)$ member is given by $\lambda_{x\,y}$. 

Applying the substitutions given by equations~\eqref{eqn_lxx}~and~\eqref{eqn_lxy} to the discretized operators~$\mat{G}^l_{\rm H}$~and~$\mat{G}^l_{\rm A}$ yields the skew-symmetric matrix

\beq
\mat{A}_{\rm H}=
 \left( \begin{array}{ccc}
0 & \zeta_z  & -\zeta_y\\
-\zeta_z  & 0 & \zeta_x\\
\zeta_y  & -\zeta_x & 0
\end{array} \right) ,
\eeq

\n and the symmetric matrix

\beq
\mat{A}_{\rm A}= \mat{b \upzeta} +\mat{\upzeta b}  -\tr(\mat{\Lambda})\mat{bb} -\bfm{b}^T\bfm{\zeta}\mat{I} ,
\eeq

\n respectively, where \mbox{$\bzeta=\mat{\Lambda}\bfm{b}$}, and $\mat{b \upzeta}$ is the dyadic formed from $\bfm{b}$ and $\bzeta$.

With these representations in place, we now look at the stability properties of various discretization schemes.\\ \\

\subsubsection{Standard discretization}

The standard discretization scheme can be written as

\beq
\bfm{B}^{l+1}=(\mat{I}-\tau\mat{G}_{\rm H}^l-\tau\mat{G}_{\rm A}^l)\bfm{B}^l .
\eeq

Inserting the numerical wave of equation~\eqref{eqn-numwave} then yields

\beq
\bfm{B}^{l+1} =(\mat{I}-\alpha r_{\rm H} \mat{A}_{\rm H} -\alpha r_{\rm A} \mat{A}_{\rm A}) \bfm{B}^{l} ,
\label{eqn_amp}
\eeq

\n where $\alpha=\tau/h^2$. 
\\ \\

\n \underline{Ambipolar diffusion}\\

\n Neglecting $\mat{A}_{\rm H}$ from equation~\eqref{eqn_amp}, the eigenvalues of the evolution operator $(\mat{I}-\alpha r_{\rm A} \mat{A}_{\rm A})$ are 

\bec
\begin{eqnarray}
          \mu_1       = && 1+\alpha r_{\rm A} \bfm{b}^T\bzeta  ,\\
          \mu_{2,\,3} = && 1+\half\alpha r_{\rm A} [ \tr(\mat{\Lambda}) \pm |\tr(\mat{\Lambda})\bfm{b}-2\bzeta|] .
\end{eqnarray}
\eec

Considering ambipolar diffusion alone, a maximum value in the eigenvalue magnitudes is found at \mbox{$\bfm{\omega}=\pi(1,\,1,\,1)$}  for an arbitrary orientation of $\bfm{B}$. The resulting stability limit is 

\beq
\tbr_{\rm A}^{\rm STD}\le\frac{1}{2}\frac{\sqrt{1+\eta^2}}{\eta} ,
\eeq

\n which is half the corresponding limit for the one-dimensional case (Paper~I).\\ \\

\n \underline{Hall diffusion}\\

\n Now neglecting $\mat{A}_{\rm A}$ from equation~\eqref{eqn_amp}, the evolution operator $(\mat{I}-\alpha r_{\rm H} \mat{A}_{\rm H})$ has eigenvalues

\bec
\begin{eqnarray}
          \mu_1 =&& 1 ,\\
\mu_{2,\,3} =&& 1\pm \mbox{i}\alpha r_{\rm H} \zeta .
\end{eqnarray}
\eec

Clearly, \mbox{$|\mu_{2,\,3}|>1$} for all $\tau>0$. The scheme therefore requires a vanishing timestep as the Hall resistivity becomes large with respect to the ambipolar resistivity such that, as in the one-dimensional case (Paper~I), 

\beq
\tbr_{\rm H}^{\rm STD}\to 0 \mbox{\hspace*{0.25cm} as \hspace*{0.25cm}} \eta\to 0 . 
\eeq

\n The standard discretization is therefore impractical for systems in which the Hall term is dominant. \\ \\

\n \underline{Mixed Diffusion} \\ 

\n Equation~\eqref{eqn_amp} does not readily allow derivation of general analytic expressions for the eigenvalues of the full amplification matrix. However, from the preceding discussions of the limiting cases where Hall and ambipolar diffusion terms are alternately neglected, and from numerical investigations of the intermediate regime, we infer a general case maximum in the magnitudes of the eigenvalues when \mbox{$\bfm{b}=(1,\,1,\,1)/\sqrt{3}$} and \mbox{$\bfm{\omega}=\omega(1,\,1,\,1)$}.  Under these assumptions the general eigenvalues of the system are

\bec
\begin{eqnarray}
          \mu_1 =&& 1-2\alpha r_{\rm A}(1-\cos\omega)^2 , \\
          \mu_{2,\,3}         =&& 1-2\alpha (r_{\rm A}\mp\mbox{i}r_{\rm H})(1-\cos\omega) (2+\cos\omega) .
	  \label{eqn-mixedeig}
\end{eqnarray}
\eec

As $\eta$ becomes small, the stability limit is dictated by $\mu_{2,\,3}$ with a maximum at \mbox{$\omega=2\pi/3$}. The corresponding timestep limit

\beq
\tbr^{\rm STD}\le\frac{8}{9}\frac{\eta}{\sqrt{1+\eta^2}} ,
\label{eqn-std}
\eeq

\n is slightly below the one dimensional limit $\eta/\sqrt{1+\eta^2}$ (Paper~I) and goes to zero with $\eta$. Again, we conclude that the standard discretization is impractical for systems in which the Hall effect is large.

\subsubsection{Super TimeStepping/Hall Diffusion Scheme}

We now present a technique for overcoming the weaknesses of the standard discretization.  Similarly to the strategy described in Paper~I, the induction equation is integrated in two parts by multiplicatively operator splitting the Hall and ambipolar terms. A technique known as Super TimeStepping (STS) is used to accelerate the timestepping for the standard discretization with ambipolar resistivity alone. However, STS does not perform well for evolution operators with complex eigenvalues and it is evident from equation~\eqref{eqn-mixedeig} that, for non-zero $r_{\rm H}$ and some orientations of $\bfm{b}$, the eigenvalues may be complex\footnote{In the one-dimensional case outlined in Paper~I, the orientation of the field was taken into account explicitly. This allowed a finite Hall diffusion term to be admitted while maintaining real eigenvalues.}. The Hall term is applied separately using a three-dimensional extension of the Hall Diffusion Scheme (HDS) introduced in Paper~I.\\ \\

\n \underline{Super TimeStepping}\\

\n STS is a technique which can be used to accelerate explicit schemes for
parabolic problems.  Essentially a Runge-Kutta-Chebyshev method, it has been 
known for some time \citep{alex}, although it remains poorly known in computational astrophysics. 

In this method a ``superstep'', $\tau^{\rm STS}$, is a composite timestep 
built up from a series of $N_{\rm STS}$ substeps such that 

\beq
\tau^{\rm STS}= \sum^{N_{\rm STS}}_{j=1} {\rm d}\tau_j .
\eeq

Optimal values for ${\rm d}\tau_j$ yield stability for the 
superstep while the normal stability restrictions on the individual 
substeps are relaxed \citep{alex}. Integrating the ambipolar diffusion term in this way yields a stability limit 

\beq
\tbr_{\rm A}^{\rm STS} =     \tbr_{\rm A}^{\rm STD} \frac{N}{2\sqrt{\nu}}\frac{(1+\sqrt{\nu})^{2N}-(1-\sqrt{\nu})^{2N}}{(1+\sqrt{\nu})^{2N}+(1-\sqrt{\nu})^{2N}} ,
\label{eqn-sts}
\eeq

\n (temporarily dropping the STS subscript from $N$ for clarity) where $\nu$ is a user-tunable damping factor and

\beq
\lim_{\nu \to 0} \tbr_{\rm A}^{\rm STS} \to N_{\rm STS}^2\tbr_{\rm A}^{\rm STD} .
\eeq

STS is first order accurate in time. In order to achieve second order accuracy Richardson extrapolation is used.\\ \\

\n \underline{Hall Diffusion Scheme}\\

\n $\mat{G}^l_{\rm H}$ is skew-symmetric and hence, dropping the ${\rm H}$ subscript for clarity,  we can write three-dimensional HDS as

\bec
\begin{eqnarray}
B_x^{l+1} &=& B_x^l - \tau(G_{x\,y}^l B_y^l +  G_{x\,z}^l B_z^l) ,
\label{HDSgen1} \\
B_y^{l+1} &=& B_y^l - \tau(G_{y\,z}^l B_z^l +  G_{y\,x}^l B_x^{l+1}) ,
\label{HDSgen2} \\
B_z^{l+1} &=& B_z^l - \tau(G_{z\,x}^l B_x^{l+1} +  G_{z\,y}^l B_y^{l+1}) .
\label{HDSgen3}
\end{eqnarray}
\eec

\n Note that equations~\eqref{HDSgen1}~to~\eqref{HDSgen3} are strictly explicit, assuming they are applied in the order shown, in the sense that all terms on the right hand sides are known. However, both equations~\eqref{HDSgen2}~and~\eqref{HDSgen3} have implicit-like terms at time $t^{l+1}$ on their right hand sides. These terms are the origin of the superior stability properties of HDS.

The order for updating the magnetic field components in equations~\eqref{HDSgen1}~to~\eqref{HDSgen3} has been arbitrarily selected. While this introduces a directional bias into the scheme, we do not find any evidence of this in the tests carried out here. Under certain conditions however, such as when there is a strong directional bias in the initial state, permutation of the order may be necessary over successive steps. We anticipate such permutation to result in a small reduction in stability however. As evidence of this, in the one-dimensional case described in Paper~I, it can easily be shown that the stable timestep limit decreases by a factor of two when the order of component updates is alternated.

In matrix form we can write three-dimensional HDS as

\beq
\bfm{B}^{l+1}=(\mat{I}-\alpha r_{\rm H} \mat{ \hat{k}\hat{k}}\mat{A}_{\rm H})(\mat{I}-\alpha r_{\rm H} \mat{\hat{\j}\hat{\j}}\mat{A}_{\rm H})(\mat{I}-\alpha r_{\rm H} \mat{\hat{\i}\hat{\i}}\mat{A}_{\rm H}) \bfm{B}^{l} ,
\label{eqn-hds}
\eeq

\n where $\mat{\hat{\i}\hat{\i}}$, $\mat{\hat{\j}\hat{\j}}$, and $\mat{\hat{k}\hat{k}}$ are dyadics formed from the unit vectors $\hat{\bfm{\i}}$, $\hat{\bfm{\j}}$, $\hat{\bfm{k}}$ in the $x$, $y$, $z$ coordinate directions respectively. Then the eigenvectors of the evolution operator on the right hand side of equation~\eqref{eqn-hds} are

\bec
\begin{eqnarray}
          \mu_1 =&& 1 , \\
\mu_{2,\,3} =&& 1-\half g \pm \half\sqrt{g(g -4)} ,
\end{eqnarray}
\eec

\n where 

\beq
g=(\alpha r_{\rm H})^2 (\zeta^2 -\alpha r_{\rm H} \zeta_x\zeta_y\zeta_z) .
\eeq

\n Hence for stability we require

\beq
0\le g \le 4 .
\eeq


The most stringent restriction is obtained from \mbox{$\bfm{b}=(1/\sqrt{3})(1,\,1,\,1)$} with  \mbox{$\bfm{\omega}=(2\pi/3)(1,\,1,\,1)$} and related symmetry points. Making the appropriate substitutions, and additionally using ordinary (unaccelerated) substepping with $N_{\rm HDS}$ substeps per full timestep, we find 

\beq
\tbr_{\rm H}^{\rm HDS}\le N_{\rm HDS} \frac{4}{\sqrt{27}}\sqrt{1+\eta^2} ,
\label{eqn_tauH}
\eeq

\n which is $4/\sqrt{27}$ times the equivalent one-dimensional limit~(Paper~I). Similarly to STS, Richardson extrapolation is required to bring HDS to second order temporal accuracy. \\ \\

\n \underline{Stability of STS/HDS}\\

\n The effective stable timestep limit for the integration of both diffusion terms using STS/HDS methods may be estimated as the minimum of $\tbr_{\rm H}^{\rm HDS}$ and $\tbr_{\rm A}^{\rm STS}$ 

\beq
\tbr^{\rm STS/HDS}=
\left\{ \begin{array}{ll}
     \tbr_{\rm H}^{\rm HDS} & \mbox{ if $\eta<=\eta^*$} \\
     \tbr_{\rm A}^{\rm STS} & \mbox{ otherwise} ,
   \end{array} \right.
\eeq

\n where $\eta^*$  is the solution of \mbox{$\tbr_{\rm H}^{\rm HDS}=\tbr_{\rm A}^{\rm STS}$} and depends on the user-defined parameters $\nu$, $N_{\rm STS}$ and $N_{\rm HDS}$.

\begin{figure}
  \bec
  \leavevmode
  \psfrag{p1}[Bl][Bl][0.8][0]{\hspace*{-0.1in}$\tbr_{\rm H}$}
  \psfrag{p2}[Bl][Bl][0.8][0]{\hspace*{-0.1in}$\tbr_{\rm A}$}
  \psfrag{p3}[Bl][Bl][1.2][0]{\hspace*{-0.1in}$\eta$}
  \includegraphics[width=240pt]{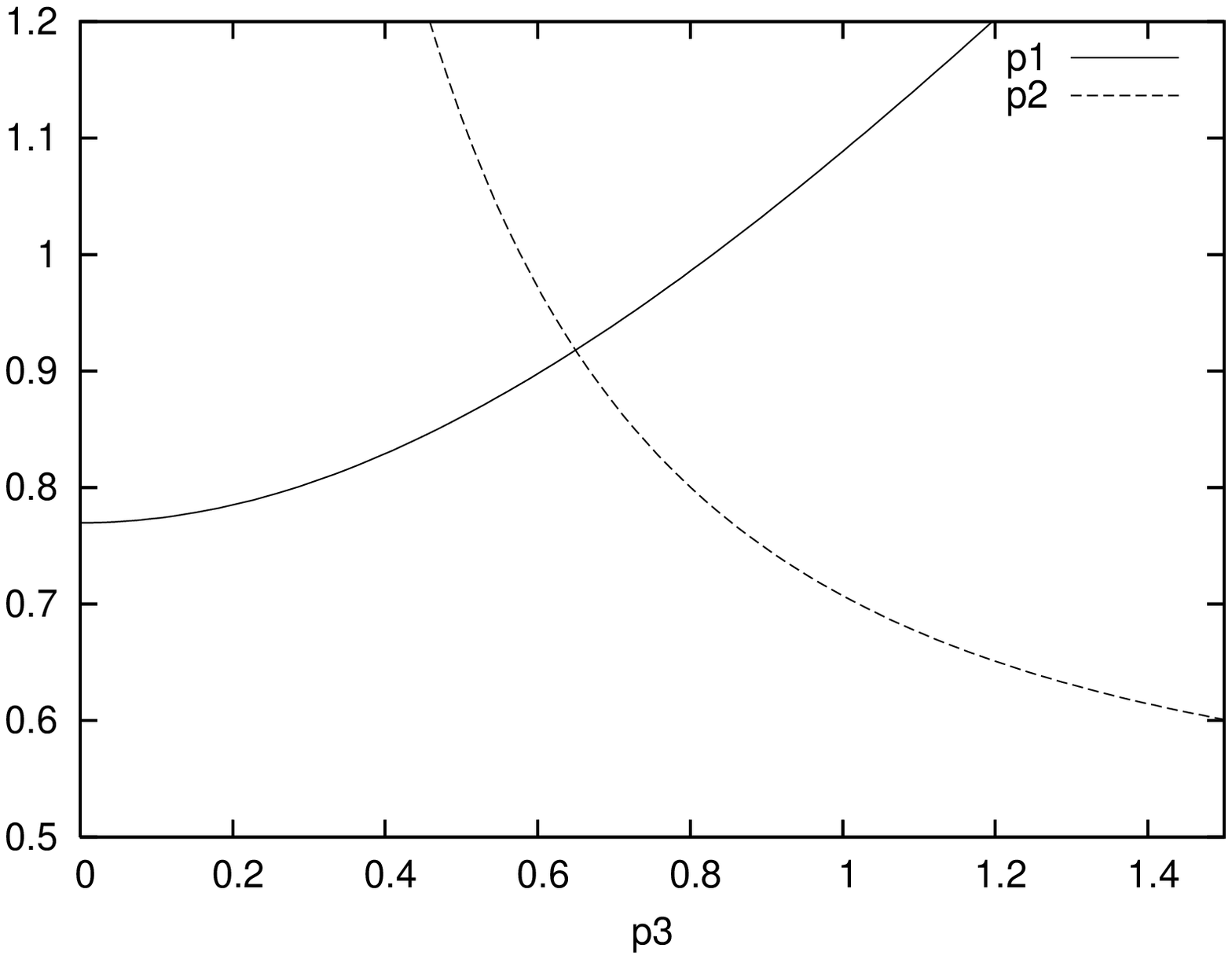}
  \caption{STS/HDS for \mbox{$\nu=0$}, \mbox{$N_{\rm STS}=1$} and \mbox{$N_{\rm HDS}=1$}: The stable timestep limits for HDS ($\tbr_{\rm H}$; solid line) and STS ($\tbr_{\rm A}$; dashed line) as functions of \mbox{$\eta\equiv r_{\rm A}/|r_{\rm H}|$}.}
  \label{fig_stshds}
  \eec
\end{figure}

In the special limiting case given by $\nu=0$, \mbox{$N_{\rm STS}=1$} and \mbox{$N_{\rm HDS}=1$}, we have \mbox{$\eta^*=\sqrt{27}/8$}. Figure~\ref{fig_stshds} illustrates that the stable timestep limit $\tbr$ has a maximum value of $\sqrt{91/108}$ at \mbox{$\eta=\eta^*$} in this case. The contrast between the maximum and minimum possible values of $\tbr$ is then only $\sqrt{91/27}$. Importantly, $\tbr$ converges to $4/\sqrt{27}$ as $\eta$ approaches zero unlike the standard scheme for which $\tbr$ goes to zero.

\subsubsection{Correction terms}
\label{sec-correction}

In the preceding sections, we considered schemes for the approximate induction equation~\eqref{B_eqn4} in the limit of small perturbations of $\bfm{B}$ about a mean field and small Ohmic resistivity $r_{\rm O}$. As previously stated, however, for consistency with equation~\eqref{B_eqn3}, the neglected small terms must be included in the scheme during each update by making the correction

\beq
\bfm{B}^{l+1} \to \bfm{B}^{l+1} +\tau ( \bfm{F}^1_{\rm O} +\bfm{F}^2_{\rm O} +\bfm{F}^2_{\rm H} +\bfm{F}^2_{\rm A}) .
\eeq

\n All terms are evaluated according to the prescriptions given by equations~\eqref{eqn-der1}~and~\eqref{eqn-der2} with the charge current $\bfm{J}$ evaluated via equation~\eqref{eqn-J}.

\subsection{$\nabla\cdot\bfm{B}=0$}
\label{sec-divb}

It is well known by now that the solenoidal condition on the magnetic field is a sensitive issue in any MHD code. In our case, however, we have found it to be particularly problematic for the tests considered here.

Both the often inaccurate Powell method \citep{powell, tothdivb} and the superior Dedner method \citep{dedner} rely on reducing the influence of numerically generated monopoles by advecting them out of the system, and also dissipating them in the case of the Dedner approach. In both cases we find that the error, while not fatal, prevents convergence in the solution at the expected rate in shock-tube tests. Additionally, when periodic boundary conditions are employed, advection cannot remove monopoles from the system and only the dissipation mechanism of the Dedner method has significant effect.

We find a variant of the constrained-transport (CT) method \citep{evanshawley}, as described in section~\ref{sec-cd}, to be effective for periodic boundary conditions but impossible to implement with fixed boundary conditions in such a way as to obtain a convergent solution for shock-tube tests. Fortunately, it is trivial to implement a projection technique in this special case as we shall discuss in section~\ref{sec-proj}.

\subsubsection{Field-interpolated centred differencing}
\label{sec-cd}

The family of CT schemes maintain $\nabla\cdot\bfm{B}$ by using the induction equation to correct the magnetic field generated by some base scheme. Usually, this has been done by constructing the electric field on a staggered mesh centred on the cell edges. \cite{tothdivb} demonstrates, however, that the staggered mesh is unnecessary if a centred differencing of the induction equation is carried out on the original grid. We make use of the field-interpolated centred differencing (field-CD) scheme he presents which has the advantage of not requiring any spatial interpolation. 

Field-CD operates by evaluating the electric field $\tilde{\bfm{E}}$ on cell centres from the base scheme using the generalized Ohm's law given by equation~\eqref{eqn-efield}. The corrected magnetic field $\bfm{B}$ is then given by a centred differencing of the induction equation

\begin{eqnarray}
B_{x\,i\,j\,k}^{l+1} & = & B_{x\,i\,j\,k}^l - \frac{\tau}{2h}\left\{
 \left(\tilde{E}_{z\,i\,j+1\,k}^{l+1}-\tilde{E}_{z\,i\,j-1\,k}^{l+1}\right)
 \right.\nonumber \\
&&-\left.\left(\tilde{E}_{y\,i\,j\,k+1}^{l+1}-\tilde{E}_{y\,i\,j\,k-1}^{l+1} 
	\right)\right\} ,
\label{eqn-cd}
\end{eqnarray}

\n and similar expressions for the remaining components of $\bfm{B}$.

In our case, since we update the magnetic field in an operator split fashion, a field-CD correction is made as each component of the electric field is applied through the base scheme. We find this is more stable than making a single correction at the end of a full update via the base scheme.

Assuming the field is initially divergence-free, equation~\eqref{eqn-cd} will conserve a centred difference definition of the magnetic field divergence

\begin{eqnarray}
(\nabla\cdot\bfm{B})_{i\,j\,k} =  
\frac{B_{x\,i+1\,j\,k}-B_{x\,i-1\,j\,k}}{2h}  \nonumber \\
+ \frac{B_{y\,i\,j+1\,k}-B_{y\,i\,j-1\,k}}{2h}+\frac{B_{z\,i\,j\,k+1}-B_{z\,i\,j\,k-1}}{2h} ,
\end{eqnarray}

\n as long as boundary conditions are compatible. Fixed boundary conditions, as required by shock tube tests, are not compatible however and an alternative approach must be taken.

\subsubsection{Projection}
\label{sec-proj}

Projection~\citep{brackbill}, similarly to CT methods, relies on a correction to the magnetic field generated by a base scheme. Briefly, the non-solenoidal component of $\bfm{B}$ is projected out of the field by solving 

\beq
\nabla^2\phi=\nabla\cdot\bfm{B} ,
\eeq

\n for $\phi$ and making the correction

\beq
\bfm{B}\to\bfm{B}-\nabla\phi .
\eeq

In Fourier space, writing \mbox{$\bfm{B}=\sum_m\bfm{B}_m$}, this amounts to projecting out the component of each mode \mbox{$\bfm{B}_m=\mbox{e}^{\mbox{i}(\bfm{\omega}_m\cdot\bfm{r})}$} parallel to the corresponding wavevector $\bfm{\omega}_m$ using

\beq
\bfm{B}_m\to\bfm{B}_m -(\hat{\bfm{\omega}}\cdot\bfm{B}_m)\hat{\bfm{\omega}} .
\label{eqn_proj}
\eeq

\section{Tests}
\label{numerical_tests}

Similarly to F03 and Paper~I, we test the numerical algorithms outlined here against the multifluid equations for weakly ionized gases in the isothermal limit with two charged species.

\begin{figure}
  \bec
  \leavevmode
  \psfrag{pp1}[Bl][Bl][1.2][-90]{\hspace*{-0.1in}$u_1$}
  \psfrag{pp2}[Bl][Bl][1.2][-0]{\hspace*{-0.1in}$x$}
  \psfrag{pp3}[Bl][Bl][1.2][-90]{\hspace*{-0.1in}$B_y$}
  \includegraphics[width=240pt]{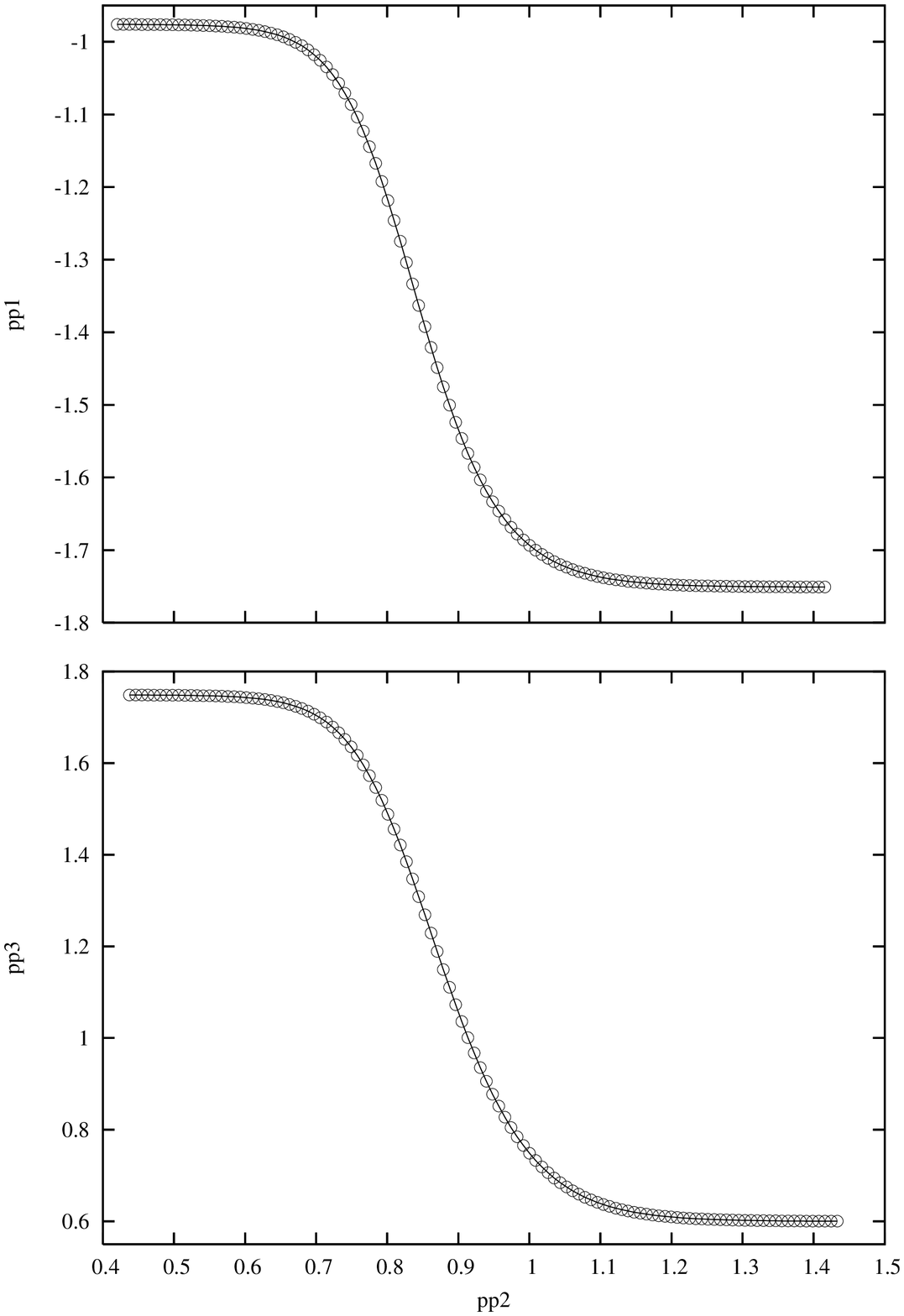}
  \caption{Neutral fluid $x$-velocity and $y$-component of magnetic field for case~A with \mbox{$h = 5 \times 10^{-3}$}.  The solution from the steady state equations, as a line, is overplotted with points from the dynamic code.}
  \label{fig_test_a}
  \eec
\end{figure}

\begin{figure}
  \bec
  \leavevmode
  \psfrag{pp1}[Bl][Bl][1.2][-90]{\hspace*{-0.1in}$u_1$}
  \psfrag{pp2}[Bl][Bl][1.2][-0]{\hspace*{-0.1in}$x$}
  \psfrag{pp3}[Bl][Bl][1.2][-90]{\hspace*{-0.1in}$B_y$}
  \includegraphics[width=240pt]{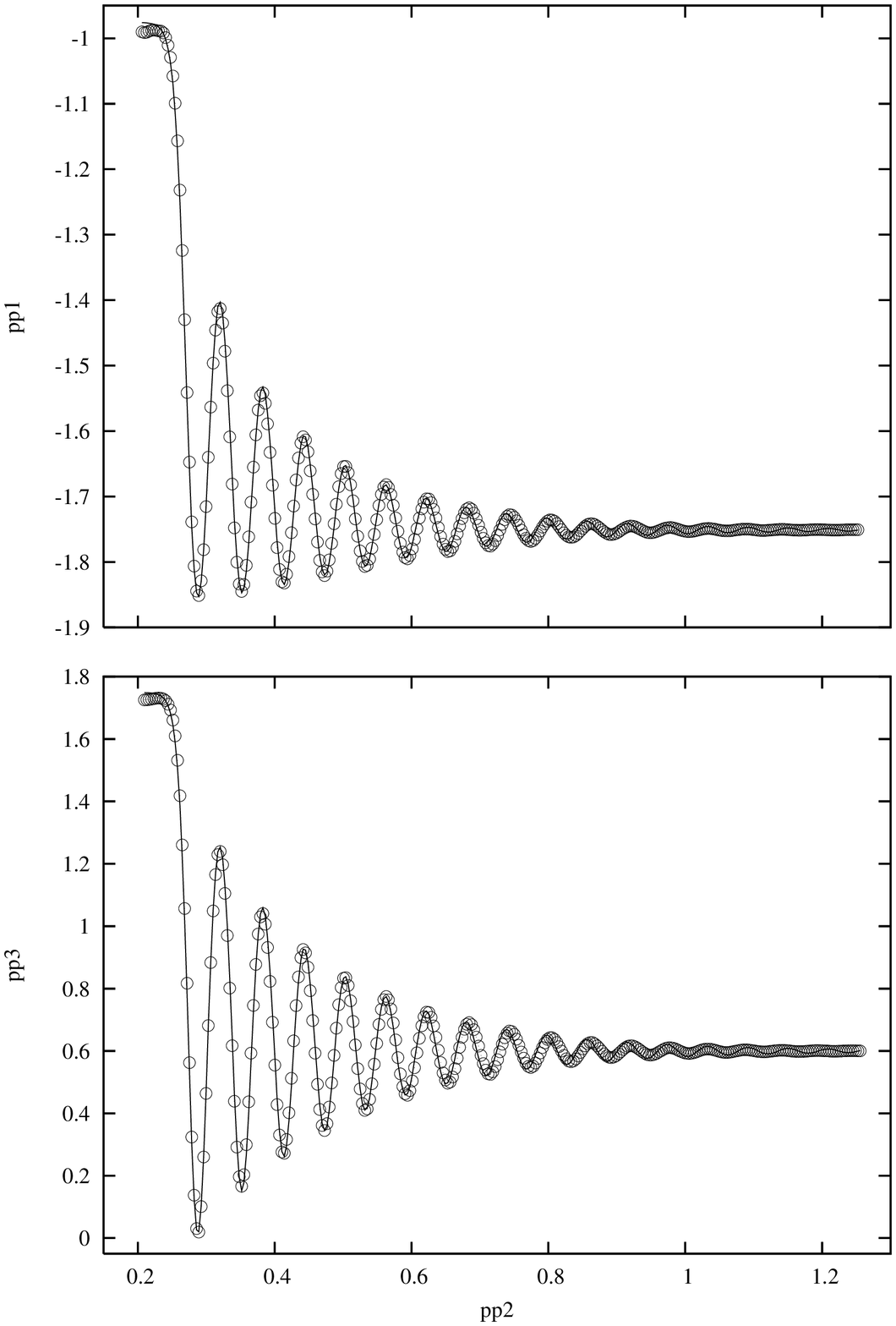}
  \caption{Neutral fluid $x$-velocity and $y$-component of magnetic field for case~B with \mbox{$h = 2 \times 10^{-3}$}.  The solution from the steady state equations, as a line, is overplotted with points from the dynamic code.}
  \label{fig_test_b}
  \eec
\end{figure}

\begin{figure}
  \bec
  \leavevmode
  \psfrag{pp1}[Bl][Bl][1.2][-90]{\hspace*{-0.1in}$u_1$}
  \psfrag{pp2}[Bl][Bl][1.2][-0]{\hspace*{-0.1in}$x$}
  \psfrag{pp3}[Bl][Bl][1.2][-90]{\hspace*{-0.1in}$B_y$}
  \includegraphics[width=240pt]{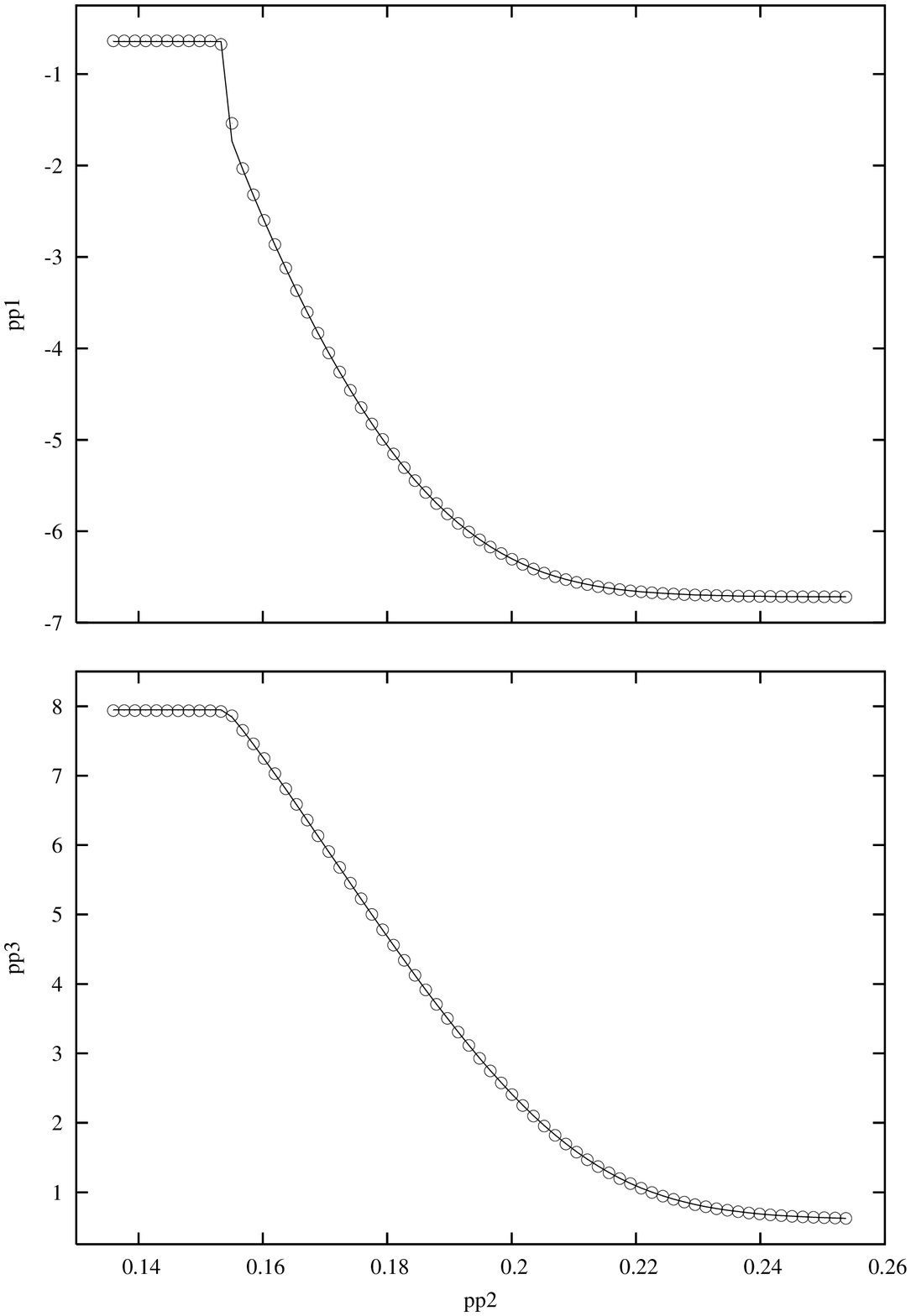}
  \caption{Neutral fluid $x$-velocity and $y$-component of magnetic field for case~C with \mbox{$h = 1 \times 10^{-3}$}.  The solution from the steady state equations, as a line, is overplotted with points from the dynamic code.}
  \label{fig_test_c}
  \eec
\end{figure}

\begin{figure}
  \bec
  \leavevmode
  \psfrag{pp3}[Bl][Bl][1.2][-90]{\hspace*{-0.1in}$u_{\mbox{\tiny{$2$}}}$}
  \psfrag{pp2}[Bl][Bl][1.2][-0]{\hspace*{-0.1in}$x$}
  \includegraphics[width=240pt]{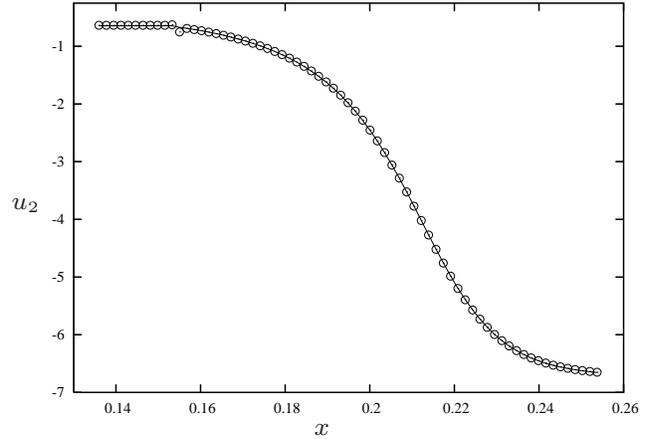}
  \caption{Negatively charged fluid $x$-velocity for case~C with \mbox{$h = 1 \times 10^{-3}$}.  The solution from the steady state equations, as a line, is overplotted with points from the dynamic code.}
  \label{fig_test_c_2}
  \eec
\end{figure}

\subsection{Shock tube tests}
\label{shocktube}

Using analytical solutions to one-dimensional problems for comparison, we run the tests obliquely to the coordinate axes in the \mbox{$(1,\,1,\,1)$} direction. An $N^3$ grid is allocated for each problem but the solution is only calculated in a narrow beam with a radius of one cell and a finite length such that it is contained completely within the grid. All cells external to the beam are referenced by their parallel displacement along the beam and treated as boundary cells. For parallel displacements outside the range of the beam, the cells are set to fixed values. Inside the beam a single reference cell is chosen at each unique value of displacement and all external cells with the same value are duplicated from this cell. In this way a properly three-dimensional problem is possible with computation only required on a small fraction of the full $N^3$ domain.

Since for this case we know that \mbox{$\hat{\bfm{\omega}}=(1,\,1,\,1)/\sqrt{3}$}, equation~\eqref{eqn_proj} simply says that for the projection method of divergence cleaning, the longitudinal component of the magnetic field must be held constant as expected trivially from the one-dimensional analogs of the solenoidal condition~\eqref{divB} and induction equation~\eqref{B_eqn}.

Similarly to F03 and Paper~I, the dynamic algorithm described here is tested against 
solutions of the steady isothermal multifluid equations.  These steady 
state equations are solved using an independent code.  The conditions for each of the tests are given in Table~\ref{test_conditions}, including the user-defined parameters $\nu$, $N_{\rm STS}$ and $N_{\rm HDS}$ for STS/HDS substepping.

\begin{table*}
\begin{tabular}{llllll} \hline
Case A & & & & & \\
Right State & $\rho_1=1$ & $\bfm{q}_1 = (-1.751,0,0)$ & $\bfm{B} =
(1,0.6,0)$ & $\rho_2=5\times10^{-8}$ & $\rho_3 = 1\times10^{-3}$ \\
Left State & $\rho_1=1.7942$ & $\bfm{q}_1 = (-0.9759,-0.6561,0)$ & $\bfm{B} =
(1,1.74885,0)$ & $\rho_2=8.9712\times10^{-8}$ & 
$\rho_3 = 1.7942\times10^{-3}$ \\
 & $\alpha_2=-2\times10^{12}$ & $\alpha_3 = 1\times 10^8$ & $K_{2\,1} = 4
\times 10^5$ & $K_{3\,1} = 2 \times 10^4$ & $a=0.1$ \\
 & $\nu = 0.05$ & $N_{\rm STS} = 5$ & $N_{\rm HDS} = 0$ & & \\
Case B & & & & & \\
Right State & As case A & & & & \\
Left State & As case A & & & & \\
 & $\alpha_2=-2\times10^{9}$ & $\alpha_3 = 1\times 10^5$ & $K_{2\,1} = 4
\times 10^2$ & $K_{3\,1} = 2.5 \times 10^6$ & $a=0.1$ \\
 & $\nu = 0$ & $N_{\rm STS} = 1$ & $N_{\rm HDS} = 8$ & & \\
Case C & & & & & \\
Right State & $\rho_1=1$ & $\bfm{q}_1 = (-6.7202,0,0)$ & $\bfm{B} =
(1,0.6,0)$ & $\rho_2=5\times10^{-8}$ & $\rho_3 = 1\times10^{-3}$ \\
Left State & $\rho_1=10.421$ & $\bfm{q}_1 = (-0.6449,-1.0934,0)$ & $\bfm{B} =
(1,7.9481,0)$ & $\rho_2=5.2104\times10^{-7}$ & 
$\rho_3 = 1.0421\times10^{-2}$ \\
 & $\alpha_2=-2\times10^{12}$ & $\alpha_3 = 1\times 10^8$ & $K_{2\,1} = 4
\times 10^5$ & $K_{3\,1} = 2 \times 10^4$ & $a=1$ \\ 
 & $\nu = 0.05$ & $N_{\rm STS} = 15$ & $N_{\rm HDS} = 0$ & & \\ \hline
\end{tabular}
\caption{Test calculation parameters.}
\label{test_conditions}
\end{table*}

\subsubsection{Case A: Ambipolar Dominated}

In this test \mbox{$r_{\rm O} = 2 \times 10^{-12}$}, \mbox{$r_{\rm H} = 1.16
\times 10^{-5}$} and \mbox{$r_{\rm A} = 0.068$} giving \mbox{$\eta = 5.86 \times 10^{3}$} and hence it can be expected that ambipolar diffusion will dominate the 
solution.  From equation~\eqref{eqn-sts} we estimate an overall speed-up of about a factor of 2 in comparison with the standard explicit approach.  Figure~\ref{fig_test_a} shows plots of the $x$ component of the neutral velocity, 
along with $B_y$ for both the dynamic and steady state solutions.  The 
calculation shown has \mbox{$h=5\times~10^{-3}$}.  Clearly the 
agreement between the two solutions is extremely good.

Since the algorithm is designed to be second order it is worthwhile
measuring the convergence rate of the dynamic solution against the solution from the steady state solver.  The comparison is made using the L1 error 
norm, $e_1$, between a section of the dynamical solution and the steady state solution.  Working from the downstream side, the section \mbox{$x_L\le x \le x_R$} is fixed about the point $x^*$ where the deviation from the downstream state first exceeds $1\%$ of the maximum variation in the solution.  Using \mbox{$x_L=x^*-0.2$} and \mbox{$x_R=x^*+0.8$} yields \mbox{$e_1=1.00\times 10^{-5}$} for \mbox{$h=5\times~10^{-3}$},
and \mbox{$e_1=9.41\times 10^{-5}$} for \mbox{$h=1\times~10^{-2}$}.  This
gives \mbox{$e_1 \propto h^{3.2}$}, above the second order convergence expected. This may be because of cross-term cancellations arising from symmetry in the \mbox{$(1,\,1,\,1)$} choice for the direction of variation in the problem.

\subsubsection{Case B: Hall Dominated}

The Hall term dominates in this test such that the overall efficiency of the scheme is governed by HDS.  The parameters chosen are \mbox{$r_{\rm O}=2\times10^{-9}$}, \mbox{$r_{\rm H}=0.0116$},
\mbox{$r_{\rm A}=5.44\times10^{-4}$} with \mbox{$\eta = 0.046 \ll 1$}\footnote{If the Hall diffusion is increased much further, it appears that the approximation of negligible charged particle inertia breaks down.}. From equations~\eqref{eqn_tauH}~and~\eqref{eqn-std} we estimate the scheme to be approximately 20 times faster than the standard explicit case.
Figure~\ref{fig_test_b} shows the results of the calculations for 
the test with \mbox{$h=2\times~10^{-3}$}.  For standard explicit codes 
the conditions lead to prohibitive restrictions on the timestep.  However, the use of HDS allows us to 
maintain a timestep close to the Courant limit imposed by the hyperbolic terms throughout the 
calculations.

As with case~A, the dynamic solution is tested to ensure it has the correct 
second order convergence characteristics.  Setting $x^*$ at the point where the solution deviates from the downstream state by $10\%$ and using \mbox{$x_L=x^*-0.05$} and \mbox{$x_R=x^*+1.0$}, we find \mbox{$e_1 = 5.11\times10^{-3}$} for \mbox{$h = 2\times10^{-3}$} and \mbox{$e_1 = 1.83\times10^{-2}$} for 
\mbox{$h = 4 \times 10^{-3}$} , giving 
\mbox{$e_1 \propto h^{1.8}$}.  The deviation from second order in this case is due to some post-shock noise in the high resolution run.

\subsubsection{Case C: Neutral subshock}

This test is similar to case~A, but with a higher soundspeed and upstream 
fast Mach number.  As a result, a subshock develops in the neutral flow because the 
interactions between the charged particles and the neutrals are not strong 
enough to completely smooth out the strong initial discontinuity in the 
neutral flow.  The ability of the algorithm described to deal with discontinuities in the solution is therefore tested. Similarly to case~A, we expect an overall speed-up of about a factor of 2 in comparison with the standard explicit approach. 

Figure~\ref{fig_test_c} shows the results of the calculations for
\mbox{$h=1\times~10^{-3}$}.  The subshock in the neutral flow is
clearly visible as a discontinuity in $u_1$, while there is no
corresponding discontinuity in $B_y$.  Figure~\ref{fig_test_c_2}
contains a plot of the $x$ component of the velocity of the negatively 
charged fluid.  There is no discontinuity in this variable,
but there are some oscillations at the point where the discontinuity in
the neutral flow occurs as already commented on by F03 and Paper~I.

It can be expected that, since there is a discontinuity in the solution of this
test and a MUSCL-type approach is used, the rate of convergence of the
dynamic solution will be close to first order, at least for resolutions high 
enough to discern the subshock in the solution.  Setting $x^*$ at the point where the solution deviates from the downstream state by $1\%$ and using \mbox{$x_L=x^*-0.02$} and \mbox{$x_R=x^*+0.1$}.  We find \mbox{$e_1 = 6.44\times10^{-3}$} for \mbox{$h = 1\times10^{-3}$} and \mbox{$e_1 = 1.16\times10^{-2}$} for \mbox{$h = 2\times10^{-3}$} yielding 
\mbox{$e_1 \propto h^{0.85}$}, close to the first order rate anticipated. As in Paper~I, we suggest the deviation from first order is due to a discontinuity in the electric field at the subshock causing an error in the charged velocities since smoothing the solution with artificial viscosity improves convergence.

\subsection{Three-dimensional MHD turbulence}

We now examine the influence of Hall and ambipolar diffusion on weakly ionized plasmas under the influence a uniform magnetic field $\bfm{B}_0$ superimposed with a weak turbulent spectrum of plane waves. \cite{wardleng99} assert that the system will evolve quite differently depending on which form of diffusion is dominant with direct consequences for molecular cloud support, angular momentum transport in accretion discs, and dynamo efficiency \citep{wardle98, wardle04, wardle99, salmeron, ss02a, ss02b, mininni05}.

\subsubsection{Initial $\bfm{B}$-field generation}

A turbulent field may be represented in a straightforward way as a sum of $M$ Fourier modes as

\beq
\bfm{B}_1(\bfm{r})=\sum^M_{m=1}A_m\mbox{e}^{\mbox{i}(\bfm{\omega}_m\cdot\bfm{r}+\beta_m)}\bfm{\hat{\xi}}_m ,
\eeq

\n where $A$, $\beta$, $\bfm{\omega}$ and $\bfm{\hat{\xi}}$ are the amplitude, phase, wave vector, and polarization vector of each mode respectively. In the limit of a continuous derivative, the solenoidal condition requires \mbox{$\bfm{\hat{\xi}}_m\cdot\bfm{\omega}_m=0$} for all values of~$m$, \ie the magnetic field is always perpendicular to the direction of propagation. 

Taking a unit cube of $100^3$ cells as the computational domain this sets a limit on the maximum allowable wavelength of \mbox{$\lambda_{max}=1/\sqrt{3}$}. Furthermore, to ensure all modes are properly resolved initially, we set the minimum wavelength $\lambda_{min}$ to 20\% of $\lambda_{max}$ such that there are more than 10 cells resolving each cycle. Logarithmic spacing in $\omega$ is then assumed such that $\Delta \omega_m/\omega_m$ is a constant where \mbox{$\Delta \omega_m\equiv \omega_{m+1}-\omega_m$}.  The amplitude $A(n)$ of each mode is generated by

\beq
A^2_m=2\sigma^2 G_m\left[\sum^M_{m=1}G_m\right]^{-1}
\eeq

\n where

\beq
G_m=\frac{\Delta V_m}{1+(\omega_m L_c)^\Gamma}
\eeq

\n The variance of the turbulent field is \mbox{$\sigma^2\equiv<B_1^2>$} through which the turbulence level $E$ is defined by \mbox{$E\equiv \sigma^2/(B_0^2+\sigma^2)$}. In the studies below we shall consider \mbox{$E=0.01$} and take the variance of the total field $<B^2>$ to be unity such that the Alfv\'enic signal speed with respect to the mean magnetic field is also unity. The correlation length $L_c$ is set to be $\lambda_{max}$ and the normalization factor $\Delta V_m$ for three-dimensional turbulence is given by

\beq
\Delta V_m=4\pi \omega_m^2\Delta \omega_m .
\eeq

\n Lastly, for a three-dimensional Kolmogorov spectrum, we use a spectral index \mbox{$\Gamma=11/3$}.

\begin{figure}
  \bec
  \leavevmode
  \includegraphics[width=240pt]{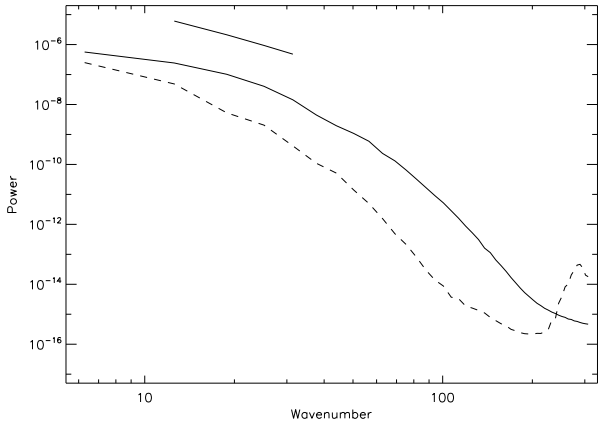}
  \caption{Density power spectra for the Hall (solid curve; \mbox{$0.653\le\rho\le 1.459$}) and ambipolar (dashed curve; \mbox{$0.891\le\rho\le 1.143$}) cases. A Kolmogorov power law (solid straight line) is also shown for reference.}
  \label{fig_rho_spectrum}
  \eec
\end{figure}


\begin{figure}
  \bec
  \leavevmode
  \includegraphics[width=240pt]{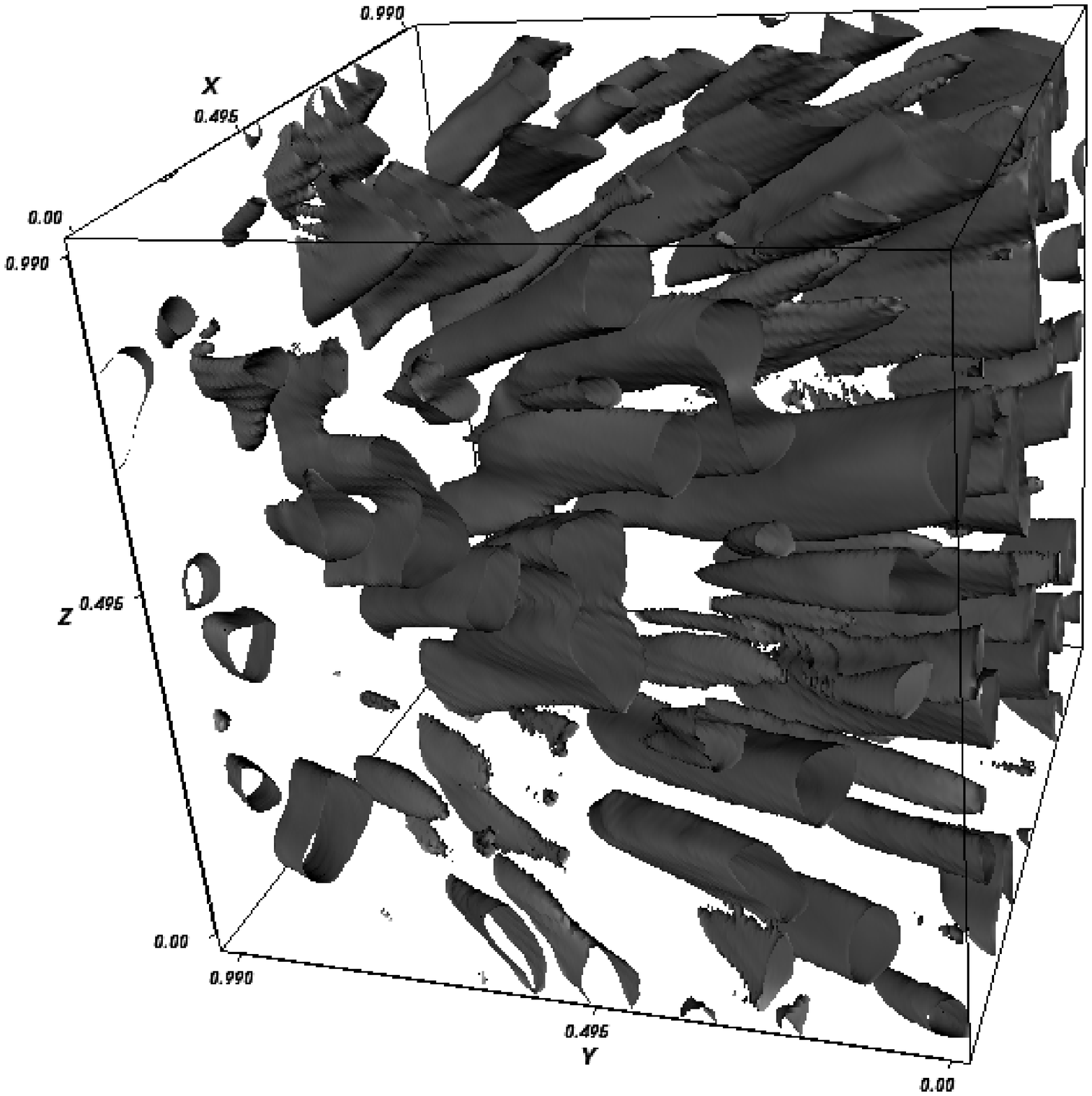}
  \caption{Ambipolar model kinetic enstrophy with isosurfaces at \mbox{$\Omega=0.12$} (\mbox{$\mx{\Omega}=0.64$} and \mbox{$<\Omega>=0.06$}).}
  \label{fig_ambi_Ek}
  \eec
\end{figure}
\begin{figure}
  \bec
  \leavevmode
  \includegraphics[width=240pt]{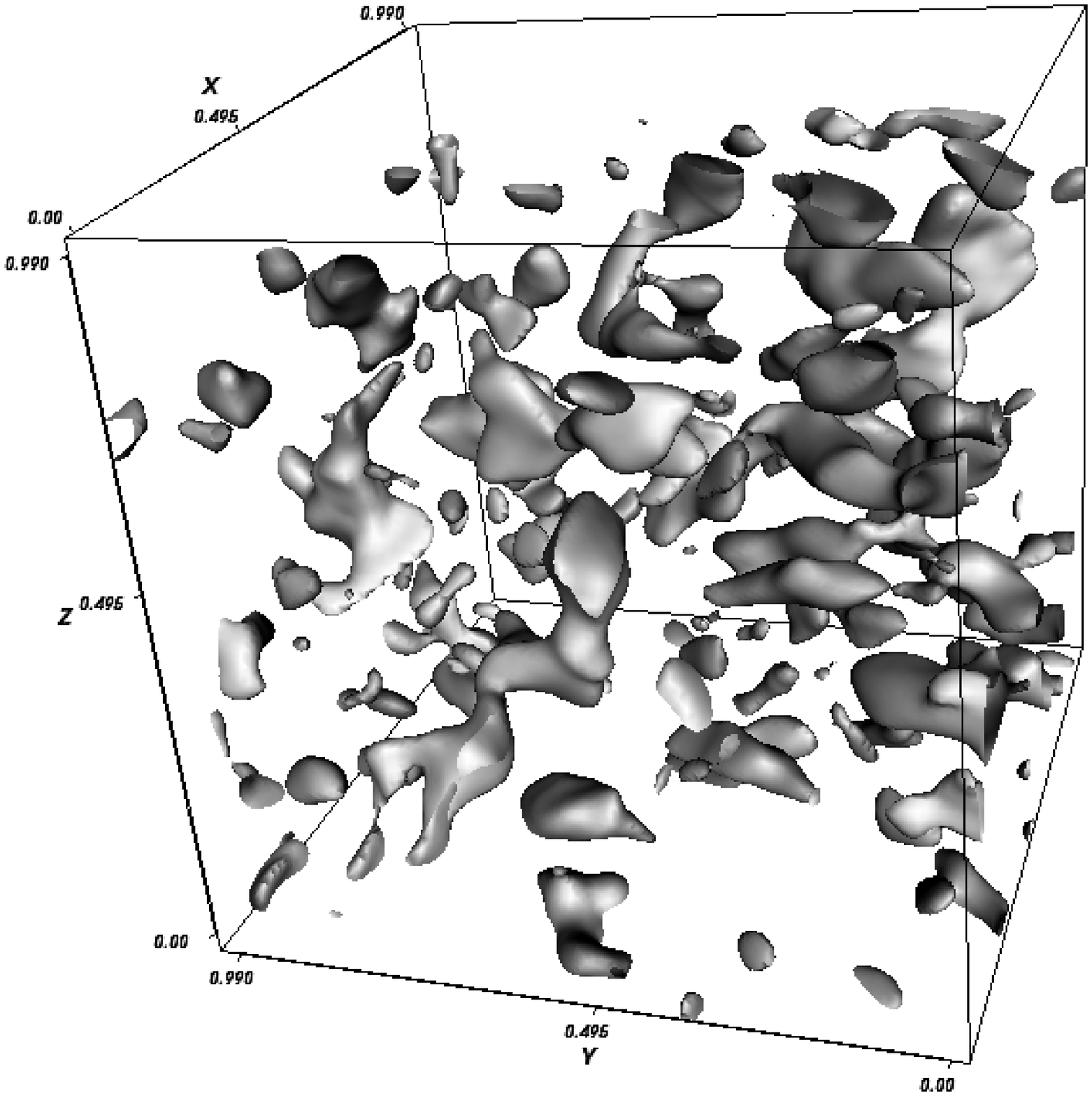}
  \caption{Hall model kinetic enstrophy with isosurfaces at \mbox{$\Omega=1.2$} (\mbox{$\mx{\Omega}=4.8$} and \mbox{$<\Omega>=0.5$}).}
  \label{fig_hall_Ek}
  \eec
\end{figure}

\subsubsection{Results}

For a first approximation to the field we use, a Mersenne twister algorithm\footnote{ See www.math.sci.hiroshima-u.ac.jp/~m-mat/MT/ewhat-is-mt.html.} is called to generate values for the phase $\beta_m$, the direction of $\bfm{\omega}_m$, and the orientation of $\bfm{\hat{\xi}}_m$ for \mbox{$M=1000$} modes. However, this field is neither divergence free nor periodic and must be modified.

Firstly, to derive a periodic field, the components of $\bfm{\omega}_m$ must be integral multiples of $2\pi$. To achieve this, the components are collapsed onto the closest lower integral multiple. Secondly, since our measure of $\nabla\cdot\bfm{B}_1$ is not continuous but discrete, the above field will not appear divergence free initially. Relaxing the condition \mbox{$\bfm{\hat{\xi}}_m\cdot\bfm{\omega}_m=0$} and assuming a centered difference approximation to $\nabla\cdot\bfm{B}_1$ on a grid of uniform spacing $h$ then yields the constraint

\beq
\bfm{\hat{\xi}}_m\cdot\sin(\bfm{\omega}_m h)=0 
\eeq

\n for a numerically divergence free field. Since $\sin(\bfm{\omega}h)$ is known explicitly, we take the polarization with respect to this quantity in order to construct an appropriate $\bfm{\hat{\xi}}$. For the particular set of modes generated for these tests, the above treatment results in $115$ unique wavevectors.

Once $\bfm{B}_1$ has been fully specified, the direction of the mean field $\bfm{B}_0$ is determined by taking a weighted average of the wavevector directions as follows

\beq
\bfm{B}_0 = B_0  \frac{ \sum^M_{m=1}A_m\bfm{\omega}_m }{ \sum^M_{m=1}\bfm{\omega}_m } .
\eeq

\n In this case we find \mbox{$\hat{\bfm{B}}_0=(0.686,\,0.608,\,-0.399)$}.

Starting from an initially uniform plasma, Fig.~\ref{fig_rho_spectrum} shows the density power spectra after 5 crossing times.  Clearly there is far more structure at all scales for the Hall case (except for some low power grid-scale noise at high frequencies). This behaviour should have significant consequences for any gravitationally unstable system.

Figure~\ref{fig_ambi_Ek} shows isosurfaces of enstrophy, defined as \mbox{$\Omega\equiv|\nabla\times\bfm{q}_1|^2$}, for the ambipolar test with isosurfaces at \mbox{$\Omega=0.12$} (\mbox{$\mx{\Omega}=0.639$} and \mbox{$<\Omega>=5.75\times 10^{-2}$}). In this case the flow has developed vortex tubes about the mean field direction. Figure~\ref{fig_hall_Ek} shows isosurfaces of enstrophy for the Hall test with isosurfaces at \mbox{$\Omega=1.2$} (\mbox{$\mx{\Omega}=4.80$} and \mbox{$<\Omega>=0.462$}). In this case the flow is more complicated showing blobs of high vorticity throughout the domain and a total enstrophy almost an order of magnitude greater than the ambipolar analog.

\begin{figure}
  \bec
  \leavevmode
  \includegraphics[width=240pt]{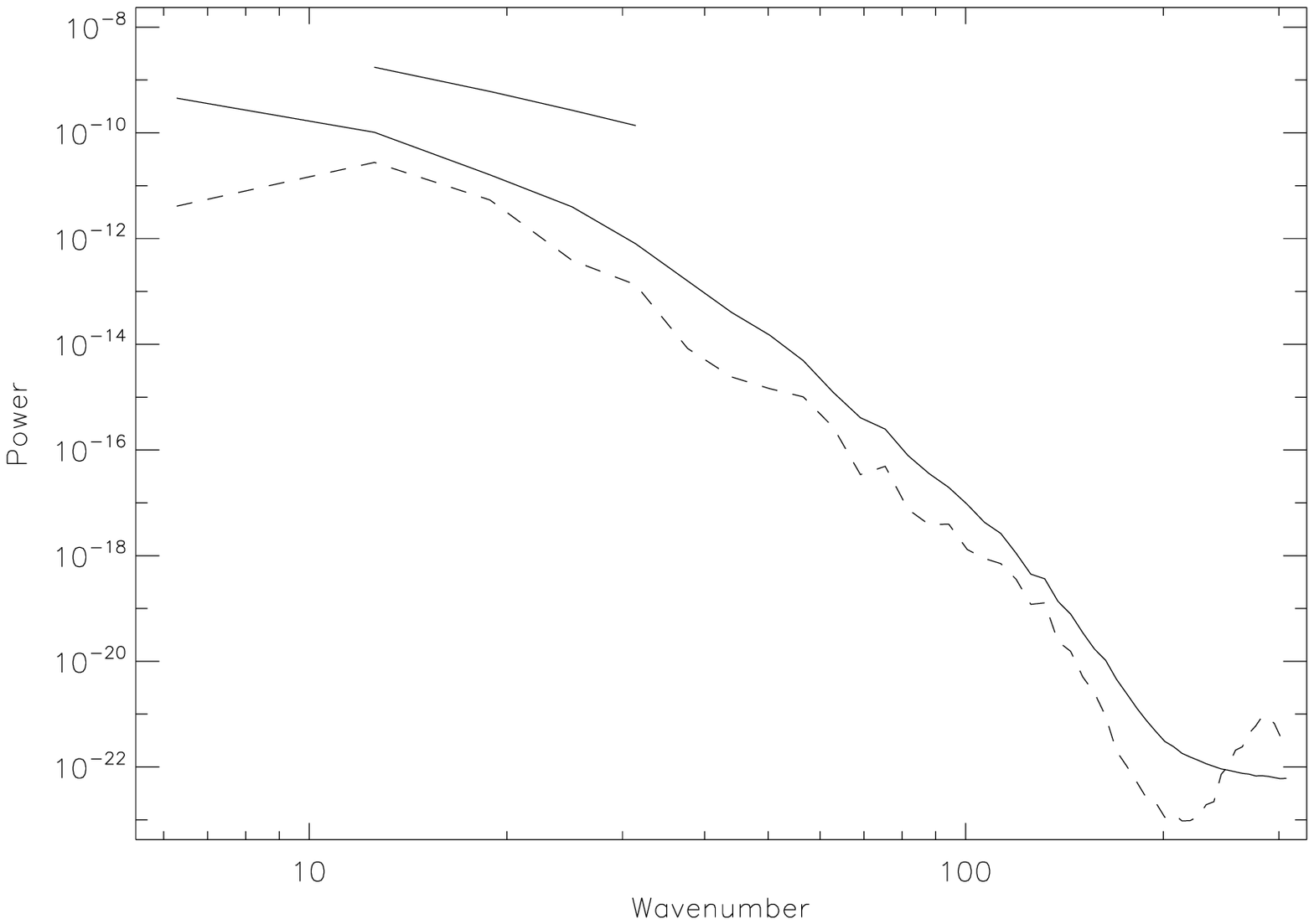}
  \caption{Helicity power spectra for the Hall (solid curve; \mbox{$\mx{|H|}=1.81\times 10^{-3}$}, \mbox{$<|H|>=3.08\times 10^{-4}$}, \mbox{$<H>=3.35\times 10^{-9}$}) and ambipolar (dashed curve; \mbox{$\mx{|H|}=6.86\times 10^{-3}$}, \mbox{$<|H|>=7.36\times 10^{-4}$}, \mbox{$<H>=-1.05\times 10^{-5}$}) cases. A Kolmogorov power law (solid straight line) is also shown for reference.}
  \label{fig_hel_spectrum}
  \eec
\end{figure}

Given its relevance to the study of dynamo action \citep[\eg][]{mininni05}, we also analyze the magnetic helicity defined by \mbox{$H\equiv \bfm{A}\cdot\bfm{B}$}, where \mbox{$\bfm{B}=\nabla\times\bfm{A}$}. Figure~\ref{fig_hel_spectrum} illustrates the power spectra for both tests. The magnetic helicity is greater at all scales for the Hall case (again, except for some low power grid-scale noise at high frequencies).  Initially, we have \mbox{$\sqrt{<H^2>}=0.0216$}, however, in the ambipolar case, by the end of the simulation the helicity has been largely dissipated to \mbox{$\sqrt{<H^2>}=0.0056$}. On the other hand, for the Hall regime test \mbox{$\sqrt{<H^2>}=0.0142$}, showing helicity is well preserved. Clearly, ambipolar and Hall diffusion have dramatically different influences on magnetic helicity.

\section{Conclusions}
\label{conclusions}

We have presented a three-dimensional numerical method for integrating the multifluid equations appropriate to weakly ionized plasmas. Crucially, the method does not rely on implicit solvers to counter the poor stability properties of conventional explicit schemes. The problematic $\nabla\times\bfm{E}'$ term describing magnetic diffusion is split into symmetric and skew-symmetric components representing ambipolar and Hall diffusion respectively (plus higher order terms). The symmetric ambipolar diffusion operator is accelerated via the Super TimeStepping (STS) method and the skew-symmetric Hall diffusion operator is treated by means of the new Hall Diffusion Scheme (HDS). A notable advantage of STS/HDS over the standard discretization is that in the limit of pure Hall diffusion, the stable timestep limit does not vanish.

Tests are presented for the special case of an isothermal three-fluid gas. For oblique shock tube problems the algorithm is accurate and converges approximately to second order when
the solution is smooth and to first order when the solution contains a
discontinuity. We also present simulations of magnetic turbulence in the ambipolar and Hall regimes and find that the evolution of the gas is very different in each case. This result may have profound implications for environments such as dense molecular clouds where magnetic turbulence is important in supporting the cloud against gravitational collapse as well as facilitating the formation of dense cores.

The local nature of the explicit scheme means it is straightforward to extend to a parallelised AMR context. This is in contrast to implicit methods for which this extension is difficult.

\section*{Acknowledgments}          

This work was partly funded by the CosmoGrid project, funded under the
Programme for Research in Third Level Institutions (PRTLI) administered by the 
Irish Higher Education Authority under the National Development Plan and with 
partial support from the European Regional Development Fund.

The authors are grateful to the School of Cosmic Physics at the Dublin
Institute for Advances Studies for facilitating this collaboration. SOS thanks David Golden for assistance with the CosmoGrid UCD Rowan cluster, Brian Reville for VTK scripts, and Sarah Tanner for helpful discussions. The authors thank the referee, Sam Falle, for insightful comments and suggestions.

\appendix

\section{Charged velocities}
\label{charged_vel_app}

For this work the collisional coefficients $K_{n\,1}$ are assumed to be 
independent of velocities and temperatures.  The following derivation (S.A.E.G. Falle, private communication) is a simplified version of the procedure outlined in Paper~I.

Transforming to the frame co-moving with the neutral gas, equation~\eqref{charged_mom} can be written as

\beq
\bfm{q}'_n \times \bfm{B} -\kappa_n\bfm{q}'_n = -\bfm{E}'
\eeq

\noindent where \mbox{$\kappa_n\equiv\rho_1 K_{n\,1}/\alpha_n$} and $\bfm{E}'$ may be derived from equations~\eqref{eqn_E1}~through~\eqref{eqn-hallpar} and equation~\eqref{eqn-J}.

The solutions for the charged species' velocities are given by

\beq
\bfm{q}'_n=-\bfm{A}^{-1}_n\bfm{E}'
\eeq

\noindent where

\beq
\bfm{A}_n = \left(\begin{array}{ccc}

-\kappa_n & B_z & -B_y \\
-B_z & -\kappa_n & B_x \\
B_y & -B_x & -\kappa_n 

\end{array}\right) .
\eeq

As in Paper~I, this procedure must be carried out iteratively if the collisional coefficients $K_{n\,1}$ are in fact dependent on the velocities of the charged species. If also required for $K_{n\,1}$, equation~\eqref{charged_en} may be used to derive the temperatures.

We point out that interpolating the primitive quantities to the cell edges before calculating the charged velocities achieves smoother results than by calculating the velocities at the cell centres and subsequently interpolating to the edges.  

\label{lastpage}


\begin{thebibliography}{99}

\bibitem[\protect\citeauthoryear{Alexiades, Amiez \& Gremaud}{1996}]{alex} Alexiades V., Amiez G., Gremaud P., 1996, Com. Num. Meth. Eng., 12, 31

\bibitem[\protect\citeauthoryear{Brackbill \& Barnes}{1980}]{brackbill} Brackbill J.U., Barnes D.C., 1980, JCP, 35, 426


\bibitem[\protect\citeauthoryear{Ciolek \& Roberge}{2002}]{cr02} Ciolek G.E., Roberge W.G., 2002, ApJ, 567, 947

\bibitem[\protect\citeauthoryear{Cowling}{1956}]{cowling} Cowling T.G., 1956, MNRAS, 116, 114


\bibitem[\protect\citeauthoryear{Dedner \etal}{2002}]{dedner} Dedner A., Kemm F., Kr\"{o}ner D., Munz C.-D., Schnitzer T., Wesenberg M., 2002, JCP, 175, 645

\bibitem[\protect\citeauthoryear{Draine}{1980}]{draine80} Draine B.T., 1980, ApJ, 241, 1021

\bibitem[\protect\citeauthoryear{Evans \& Hawley}{1988}]{evanshawley} Evans C.R., Hawley J.F., 1988, ApJ, 332, 659


\bibitem[\protect\citeauthoryear{Falle}{2003, hereafter F03}]{falle03} Falle S.A.E.G., 2003, MNRAS, 344, 1210 (F03)


2004, ApSS, 293, 83 



\bibitem[\protect\citeauthoryear{Huba}{2005}]{huba} Huba J.D., 2005, Proc. ISSS-7, Kyoto Univ. \\(http://www.rish.kyoto-u.ac.jp/isss7/CDROM/\\CONTENTS/DATA\_PDF/T-JHUB.PDF)


\bibitem[\protect\citeauthoryear{Mininni, G\'{o}mez \& Mahajan}{2005}]{mininni05} Mininni P.D., G\'{o}mez D.O., Mahajan S.M., 2005, ApJ, 619, 1019 

\bibitem[\protect\citeauthoryear{Mestel \& Spitzer}{1956}]{mestel56} Mestel L., Spitzer Jr. L., 1956, MNRAS, 116, 50 

\bibitem[\protect\citeauthoryear{O'Sullivan \& Downes}{2006, hereafter Paper~I}]{osd06} O'Sullivan S., Downes T. P., 2006, MNRAS, 366, 1329 (Paper~I)


\bibitem[\protect\citeauthoryear{Pandey \& Wardle}{2006}]{pandey} Pandey B.P., Wardle M., 2006, preprint (astro-ph/0608008)

\bibitem[\protect\citeauthoryear{Powell}{1994}]{powell} Powell K.G., 1994, ICASE Report No. 94-24, Langley, VA

\bibitem[\protect\citeauthoryear{Salmeron \& Wardle}{2005}]{salmeron} Salmeron R., Wardle M., 2005, MNRAS, 361, 45

\bibitem[\protect\citeauthoryear{Sano \& Stone}{2002a}]{ss02a} Sano T., Stone J.M., 2002a, ApJ, 570, 314

\bibitem[\protect\citeauthoryear{Sano \& Stone}{2002b}]{ss02b} Sano T., Stone J.M., 2002b, ApJ, 577, 534


\bibitem[\protect\citeauthoryear{Smith \& Mac Low}{1997}]{sm97} Smith M.D., Mac Low, M.-M., 1997, A\&A, 326, 801

\bibitem[\protect\citeauthoryear{Spitzer}{1956}]{spitzer} Spitzer Jr. L., 1956, Physics of Fully Ionized Gases. Interscience Publishers, Inc., New York

\bibitem[\protect\citeauthoryear{Spitzer}{1978}]{spitzer78} Spitzer L., 1978, Physical Processes in the Interstellar Medium. Wiley, New York

\bibitem[\protect\citeauthoryear{Stone}{1997}]{stone97} Stone J.M., 1997, ApJ, 487, 271

\bibitem[\protect\citeauthoryear{Strang}{1968}]{strang} Strang G., 1968, SIAM J. Numer. Anal., 5, 505

\bibitem[\protect\citeauthoryear{T\'oth}{1994}]{toth94} T\'oth G., 1994, ApJ, 425, 171

\bibitem[\protect\citeauthoryear{T\'oth}{2000}]{tothdivb} T\'oth G., 2000, JCP, 161, 605

\bibitem[\protect\citeauthoryear{Wardle}{1998}]{wardle98} Wardle M., 1998, MNRAS, 298, 507

\bibitem[\protect\citeauthoryear{Wardle}{1999}]{wardle99} Wardle M., 1999, MNRAS, 307, 849

\bibitem[\protect\citeauthoryear{Wardle \& Ng}{1999}]{wardleng99} Wardle M., Ng C., 1999, MNRAS, 303, 239

\bibitem[\protect\citeauthoryear{Wardle}{2004}]{wardle04} Wardle M., 2004, ApSS, 292, 317

\end{thebibliography}
\end{document}